\documentclass[sn-chicago]{sn-jnl}

\usepackage{graphicx}%
\usepackage[shortlabels]{enumitem}%
\usepackage{multirow}%
\usepackage{amsmath,amssymb,amsfonts}%
\usepackage{amsthm}%
\usepackage{mathrsfs}%
\usepackage{braket}%
\usepackage[title]{appendix}%
\usepackage{xcolor}%
\usepackage{textcomp}%
\usepackage{manyfoot}%
\usepackage{booktabs}%
\usepackage{algorithm}%
\usepackage{algorithmicx}%
\usepackage{algpseudocode}%
\usepackage{listings}%
\usepackage{comment}

\theoremstyle{thmstyleone}%
\newtheorem{theorem}{Theorem}
\theoremstyle{thmstyletwo}%

\theoremstyle{thmstylethree}%
\newtheorem{definition}{Definition}%

\begin{document}

\title[Article Title]{Symmetry Breaking through Superselection by Boundary Conditions}

\author*[1]{\fnm{Silvester G.A.} \sur{Borsboom}}\email{silvester.borsboom@ru.nl}

\author*[2]{\fnm{José P.} \sur{Dupont}}\email{dupont@physics.leidenuniv.nl}

\affil[1]{\orgdiv{Institute for Mathematics, Astrophysics and Particle Physics and Radboud Center for Natural Philosophy}, \orgname{Radboud University}, \orgaddress{
\city{Nijmegen}, 
\country{The Netherlands}}}

\affil[2]{\orgdiv{Institute-Lorentz for Theoretical Physics and Leiden Institute of Physics}, \orgname{Leiden University}, \orgaddress{
\city{Leiden}, 
\country{The Netherlands}}}

\abstract{Spontaneous symmetry breaking (SSB) is central to modern physics but is conventionally defined only for infinite systems, raising challenges for its interpretation in finite, real-world setups. This paper argues that the key to resolving this issue lies in the underappreciated role of \textit{boundary conditions} in quantum systems.
Inspired by both the relational approach to symmetries and the physical mechanism behind symmetry breaking, we formulate a relational interpretation of SSB: a finite system exhibits SSB relative to a reference environment which can induce perturbations across the boundary.
This eliminates the need for the thermodynamic limit, offering a more physical picture of SSB that emphasizes the observable consequences of the interactions that real-life systems inevitably have with their environment. 
We show how, in this relational interpretation, SSB for both lattice systems and (gauge) field theories should be understood as subtle, rather than spontaneous, symmetry breaking, still in contrast to explicit symmetry breaking.
We also explain how algebraic definitions of SSB for infinite systems relate to the intuitive picture of SSB in finite systems and illustrate how asymptotic boundary conditions push the environment ``to infinity''.
In this way, our relational interpretation of SSB provides a unified conceptual framework applicable to symmetry-breaking in systems of any size.}




\maketitle

\newpage
\tableofcontents

\section{Introduction}\label{intro}

It is a commonplace to note that symmetries are ubiquitous in modern physics. It is, however, less commonplace to note that there really is much we do not understand about symmetries in physics, especially regarding the concept of \textit{symmetry breaking}. In fact, while spontaneous symmetry breaking (SSB) is relatively well understood as an equilibrium phenomenon \cite[]{anderson1952approximate, coleman1987mixed, koma1994symmetry}, multiple questions still exist about the form and even the possibility of its dynamics. Dynamical symmetry breaking is discussed for example in the context of the electroweak phase transition in cosmology \cite[]{10.21468/SciPostPhysLectNotes.24, Weir_2018} or open quantum systems and quantum computing \cite[]{Lessa_2025, hauser2026strongtoweaksymmetrybreakingopen}. More fundamentally, the mechanism that allows one to dynamically break a symmetry in quantum physics, a problem tightly connected to the measurement problem as it requires a system to evolve non-unitarily (at least effectively), is still under investigation \cite[]{vanwezel2010broken,landsmanFleaSchrodingerCat2013, landsmanFoundationsQuantumTheory2017, vanwezel2026ssb}.

A core difficulty in developing a dynamical understanding of SSB lies in the multifaceted and thereby often obscured role of boundary conditions. These facets appear in three different and often separately discussed descriptions of SSB: in terms of singular limits, by means of algebraic definitions and in (effective) field theories. The aim of this article is to relate these descriptions of SSB to one another. Interestingly, we find that these three descriptions of SSB are different not because of the number of degrees of freedom of the system under consideration, but really because of the different boundary conditions one considers.

The essential role of boundary conditions in SSB is closely related to the relational view on symmetries. Indeed, the literature on philosophy of symmetry over the past two decades has arrived at the consensus that symmetries acquire physical significance only relationally, i.e. across a boundary between subsystems \cite[]{greavesEmpiricalConsequencesSymmetries2014,tehGalileoGaugeUnderstanding2016,gomesHolismEmpiricalSignificance2021,wallaceIsolatedSystemsTheireen,wallaceIsolatedSystemstwee,tehPhilosophySymmetry2024}. This idea is often referred to as \textit{Galileo's ship}, but it has not yet been applied systematically to SSB. If one subscribes to this relational view, then symmetry breaking must also be viewed through a relational lens: one can tell how a system breaks a symmetry only in relation to an external reference frame, separated from the system by a boundary or some other kind of separation, which could simply be the fact that the system actually ends somewhere. Any real-life system, however, is not perfectly isolated, enabling the possibility of SSB occurring in this system via interaction with its environment through its boundary. We believe that this viewpoint can shed light on SSB in both finite and idealized infinite systems. Indeed, one reason that SSB in finite systems has been difficult to relate to SSB in infinite systems and their algebraic descriptions is that infinite systems seem to take some sort of a \textit{view from nowhere}, where the environment is located at infinity through \textit{asymptotic} boundary conditions, whereas real-life finite systems are clearly viewed from somewhere (say, the lab) and are separated from their environment by a real, finite boundary. Incorporating the role of asymptotic and finite boundaries for the physical significance of symmetries can therefore improve our understanding of SSB.

This boundary-centered perspective has a clear antecedent in the rigorous treatment of classical spin systems. There, infinite-volume equilibrium states are not defined by a single Boltzmann factor for the whole infinite lattice, but by Gibbs measures: probability measures whose conditional distribution on every finite region is a finite-volume Gibbs distribution with the exterior configuration held fixed as a boundary condition \cite[]{Dobrushin1968,LanfordRuelle1969,Georgii}. At low temperature, the same symmetric interaction can admit several infinite-volume Gibbs measures, obtained for instance as limits with plus or minus boundary conditions in the Ising model. These extremal Gibbs measures represent distinct pure phases and break the spin-flip symmetry, while their symmetric mixture is non-extremal. Thus, in classical spin systems, the dependence of SSB on boundary conditions is already encoded in the Gibbs-measure formalism. Our aim is to extract this lesson and apply it more broadly, including to quantum systems and field theories (both classical and quantum), while also framing it explicitly in the relational language of the modern philosophical literature on symmetry.

The interplay between boundaries and symmetries has recently received much attention from physicists and philosophers alike, often in the context of gauge theories,\footnote{See e.g. \cite[]{Rovelli:2013fga,Donnelly:2016auv,gomesUnifiedGeometricFramework2019,gomesGaugingBoundaryFieldspace2019,gomesHolismEmpiricalSignificance2021,gomesQuasilocalDegreesFreedom2021,rielloHamiltonianGaugeTheory2024,Borsboom:2025agn,borsboomdeharo,borsboominstantaneous}.} but also more generally in relation to (quantum) reference frames \cite[]{Kabel:2024lzr}. In this article we defend the perspective that the interplay between boundaries and symmetries is not just a gauge-theoretical issue, but also plays a central role in any model with a global symmetry, such as the Ising model of a ferromagnet. In gauge theories boundaries are particularly challenging because of the additional need to quotient out unphysical (gauge) degrees of freedom. However, symmetry breaking in gauge and non-gauge theories relies on a common idea: both gauge and non-gauge symmetries can become physical only across a boundary, namely when the symmetry transformation acts differently on a subsystem of interest than on its environment. Thus we espouse a Leibnizian view, in which there are (contra Newton) no absolute but only relative reference frames.

Depending on the nature of the system and its boundary, different types of boundary conditions can be chosen. We argue that asymptotic boundary conditions on infinite quantum systems, which lead to superselection sectors, simulate a system in perfect isolation. For finite quantum and classical systems, a Dirichlet type of boundary condition may represent an actual neighboring object or end to the system, and can lead to something akin to a superselection structure, with the sectors labeled by the allowed boundary states. As we will see, it is natural to represent the influence of such a neighboring object or external environment via a perturbation in the system's Hamiltonian, forcing the system into one of its stable, symmetry-broken sectors. In this spirit, we develop a novel view of symmetry breaking for both infinite and finite, gauge and non-gauge systems that is centered around the idea of \textit{superselection through boundaries}. Our approach might be viewed as a boundary-centered development of Wallace's statement that ``the real signature of SSB in finite systems is the breakdown of those systems' state spaces [...] into sectors that are approximately dynamically isolated'' \cite[]{Wallace:2018zbu}, where the boundary selects in which symmetry-broken sector the system is found.

To develop our proposal we commence with an exploration of existing explanations of SSB in Section~\ref{sec.defining_ssb}, using the transverse-field Ising model and the BCS model of superconductivity as case studies. We highlight how the co-existence of long-range order and a global symmetry can lead to perturbation-induced symmetry breaking in the thermodynamic limit, provided that the system is not isolated and interacts with another system via its boundary. We argue that this makes SSB in real-life finite systems only \textit{seemingly} spontaneous, and that it is therefore better called \textit{subtle} symmetry breaking. Then, in Section~\ref{secthermoidealization}, we discuss the algebraic treatment of SSB, suitable to describe infinitely large systems. Here we relate the algebraic formalism to SSB in finite systems, and point out the implicit role of asymptotic boundaries.
In Section~\ref{relationalsymmetries} we turn to the Abelian Higgs model as an exemplar of SSB in both high-energy physics and, as an effective field theory, in condensed matter physics. Here, the story is more nuanced due to the breaking of a gauge (instead of a non-gauge) symmetry. We show how the gauge symmetry of a field theory on Minkowski spacetime becomes physical through an appropriate choice of asymptotic boundary conditions, including in particular fall-off conditions on the order parameter. We then argue that boundary conditions on the order parameter must similarly be implemented if the field theory is defined on a finite region of spacetime, as is the case for e.g. a finite-size superconductor. Section~\ref{conclusion} concludes: by paying close attention to the hitherto implicit role of boundary conditions, we show how SSB can be uniformly understood as a system being compelled into a specific symmetry-broken state or superselection sector due to influence through the boundary. 

\section{From Spontaneous to Subtle Symmetry Breaking}\label{sec.defining_ssb}

In this Section we explore the mechanism and defining properties of SSB, i.e. the phenomenon in which the ground or thermal equilibrium state of a large system breaks the symmetry of its dynamics (governed by a Hamiltonian or Lagrangian), seemingly spontaneously.
We explain how in equilibrium SSB can be understood for systems of both infinite and finite size. We distinguish between two lattice models, namely the transverse-field Ising model with a \textit{discrete} global symmetry and the BCS theory of superconductivity with a \textit{continuous} global (gauge) symmetry and stress the key role that boundaries play in selecting a symmetry-broken ground state for these models.

\subsection{The textbook definition}\label{sec.textbookdefn}

It is useful to start our discussion with a textbook definition of SSB, which aims to describe the phenomenon in a somewhat formal way.
Here, and in the rest of this Section, we consider a typical quantum system on a $d$-dimensional lattice. The system is of finite size, i.e. we consider a subset of the lattice of $N = n^d<\infty$ sites. The dynamics of the system is governed by a Hamiltonian $H_N \in B(\mathcal{H}_N)$, where $\mathcal{H}_N$ is the Hilbert space of the total system. The Hamiltonian is \textit{local}, meaning that it only contains interactions between $i$th nearest-neighbors, with $i \leq r_c$ an $N$-independent parameter.\footnote{There exist models with a nonlocal Hamiltonian, such as the Lieb-Mattis model of anti-ferromagnetism, that exhibit SSB. We focus on local lattice models simply because our main study objects, the transverse-field Ising model and the BCS model, have a local Hamiltonian and together offer rich enough behavior for our discussion.}
For example, the local Hamiltonian $H_N \in \mathcal{A}_N:=B(\mathbb{C}^{2^N})$ of the \textit{transverse-field Ising model} is defined by \cite[]{schultz1964two, pfeuty1970one, kogut1979introduction}
\begin{equation}\label{eq.local_ising_hamiltonian}
    H_{N} = -\sum_{x\in\Lambda}\sigma_3(x)\sigma_3(x+1) + B\sum_{x\in\Lambda} \sigma_1(x).
\end{equation}
Here, $\sigma_{i}(x)$ is the $i$th Pauli spin matrix applied to site $x$ of a one-dimensional lattice $\Lambda \subset \mathbb{Z}$, $|\Lambda|=N$, $B$ is the transverse field strength and the spin coupling is ferromagnetic. The model has only nearest-neighbor interactions, so $r_c=1$.

We say a system possesses a \textit{symmetry group} $G$ if the Hamiltonian is invariant under the action of any transformation $g \in G$. This means the unitary operator $U(g)$ representing the transformation commutes with the Hamiltonian: $[H_N, U(g)] = 0$ for all $g \in G$. For example, the transverse-field Ising Hamiltonian has symmetry group $\mathbb{Z}_2$ generated by the global spin-flip operator
\begin{equation}\label{eq.z2_symmetry_ising}
    U_N = \prod_{x\in\Lambda} \sigma_1(x).
\end{equation}
SSB can now be defined as a property of a stable state, commonly meaning a ground or thermal equilibrium state of the system, that is therefore invariant under time evolution \cite[]{vanwezel2026ssb}.\footnote{A similar definition exists for systems described by a Lagrangian or action.}
\\
\begin{definition}\label{def.ssb_textbook_finite}
    Consider a system with a Hamiltonian $H_N$ defined on a Hilbert space $\mathcal{H}_N$, and with symmetry group $G$. Let $\rho: \mathcal{H}_N\to\mathcal{H}_N$ be a stable state of the system.
    Then $G$ is said to be spontaneously broken by $\rho$ if:\footnote{For a pure state $\rho = \ket{\psi}\bra{\psi}$ the definition reduces to $U(g)\ket{\psi} \neq e^{i\varphi}\ket{\psi}$ for any $\varphi \in \mathbb{R}$.}
    \begin{equation}\label{eq.ssb_finite}
        \exists g\in G: U(g)\rho U(g)^* \neq \rho.
    \end{equation}
\end{definition}

\noindent According to this definition, when SSB occurs the system is found in a state that is invariant under time evolution and does not exhibit the symmetries of the Hamiltonian under which the system evolves, i.e. there is a mismatch between the symmetries of the system's fundamental laws and the symmetries of its observed state. This is surprising, for what causes the system to choose a symmetry-broken state when the dynamics are perfectly symmetric? The system seems to have \textit{spontaneously}\footnote{For an in-depth study of the meaning of the term `spontaneous' in natural philosophy and physics, including in relation to SSB, see \cite{borsboomspontaneity}.} (i.e. without a cause) chosen to be in one of many symmetry-broken states.

However, in our view, the above notion of stability leads to confusion both about the possibility of SSB and the precise condition leading to SSB. First of all, most finite systems whose thermodynamic limit would satisfy Definition~\ref{def.ssb_textbook_finite} do not satisfy it themselves, since their symmetry-broken states are no ground states and often not even energy eigenstates. This raises the question of how finite, real-life systems can spontaneously break a symmetry. It is possible to answer this question, and we will do so in Section~\ref{ssec.mechanism}, but only if one accepts a different, more physical notion of stability. 
Secondly, there are non-interacting systems that possess symmetry-broken ground states which do satisfy Definition~\ref{def.ssb_textbook_finite}, but are very different from the systems that are generally considered to \textit{spontaneously} break a symmetry.
Consider for example a one-dimensional chain of non-interacting spin-1/2 particles. The Hamiltonian $H$ of this system is equal to zero and trivially has a global spin-flip symmetry of e.g. $\prod_x \sigma_1(x)$. Any configuration of spins will be a ground state of the system, but e.g. the ground state $\ket{000\dots0}$ breaks the global spin-flip symmetry of $H$, thus satisfying Definition~\ref{def.ssb_textbook_finite}. However, any local perturbation $\epsilon \sigma_3(x)$ of the Hamiltonian will locally change the ground state configuration to a different symmetry-broken state. In other words, the symmetry-broken ground state, while invariant under time-evolution, is not physically stable in any real-life situation. As a consequence, if perturbations come from all sides, on average the system will be symmetric. But because of the lack of stability, this form of symmetry-breaking is not persistent, unlike the phenomena which SSB is meant to describe.

Therefore, we introduce a second notion of stability, based on whether the state would be stable in a real-life setting. 
This notion is twofold. First of all, states need to be stable under local measurements, meaning that measuring e.g. the magnetization at position~$x$ does not affect the outcome of a later measurement at far-away position~$y$. This is equivalent to satisfying the cluster decomposition~\cite[]{vanwezel2026ssb}, which states that for any two local observables $a(x), b(y)$
\begin{equation}
    C_{ab}(x,y) := \braket{a(x)b(y)} - \braket{a(x)}\braket{b(y)} \to 0 \text{ for } |x-y|\to\infty.
\end{equation}
Secondly, the state has to be invariant under local perturbations of the Hamiltonian to avoid the symmetry-broken state evolving simply to a different state. This \textit{dynamical stability} is guaranteed by the state having (quasi) long-range order. A state $\ket{\psi}$ is said to have \textit{quasi long-range order} if
\begin{equation}
    \braket{O(x)^{\dagger}O(y)} := \braket{\psi|O(x)^{\dagger}O(y)|\psi} \propto \frac{1}{|x-y|^{\alpha}}, \quad \alpha 
    > 0,
\end{equation}
i.e. polynomial instead of exponential decay of correlations in the local order parameter $O(x)$. If $\alpha=0$, the correlation length diverges and the state is said to have \textit{long-range order} \cite[]{vanwezel2026ssb}. 
When all symmetry-broken states have long-range order, tunneling from one state to another is exponentially suppressed. Different symmetry-broken states have a different value of their order parameter, and because of their long-range order differ at order $N$ sites. This means that tunneling from one state to another scales as $\exp(-N)$. For typical systems of $N \approx 10^{23}$ particles, $\exp(-N) \approx 0$, meaning that the symmetry-broken state is stable with respect to local perturbations of the Hamiltonian. In the thermodynamic limit $N\to\infty$ tunneling rates become exactly zero, leading to superselection sectors, which we will treat in Section~\ref{secthermoidealization}.

The description of a stable state in terms of the clustering decomposition and long range order allows for an explanation of SSB for finite systems and distinguishes SSB from accidental symmetry breaking in non-interacting systems. For this reason we take stability of states in the textbook Definition~\ref{def.ssb_textbook_finite} of SSB to refer to this second notion of stability.

\subsection{The mechanism}\label{ssec.mechanism}

In this Section we make the mechanism of SSB in finite lattice models explicit by showcasing two examples, one with a discrete and one with a continuous symmetry. We show that SSB in both cases can be understood as originating from tiny symmetry-breaking perturbations, making the symmetry-breaking only seemingly spontaneous and thus better called \textit{subtle} symmetry breaking (SSB).

Generally, the symmetry-broken states of a system with a symmetric Hamiltonian are no energy eigenstates, let alone ground states of the Hamiltonian. For example, for any $B>0, N < \infty$ the symmetry-broken all-spin-up state $\ket{\uparrow}$ of the transverse-field Ising model (recall the Hamiltonian from Eq.~\eqref{eq.local_ising_hamiltonian}) is no energy eigenstate for any $B \neq 0$ since for any $\varphi, E\in\mathbb{R}$ \cite[]{tasaki2020long}
\begin{equation}
    H_N\ket{\uparrow} = -N\ket{\uparrow}+B\sum_{x\in\Lambda} \sigma_1(x)\ket{\uparrow} \neq E e^{i\varphi}\ket{\uparrow}.
\end{equation}
This raises the question of how they can be stable states of the system. For many quantum systems the answer lies in the existence of symmetry-broken states which have approximately ground state energy and become both dynamically stable and clustering when the system approaches the thermodynamic limit (to be discussed more rigorously in Section~\ref{secthermoidealization}).

So what systems will spontaneously break the symmetry of their Hamiltonian? The answer is that a system which possesses (i) a global symmetry and (ii) only ground states with long-range order will, in real life, have an approximate ground state that spontaneously breaks this global symmetry \textit{provided that the system is large enough}. The explanation of how these approximate ground state become true ground states in the thermodynamic limit is quite technical, and we will give it first schematically and then in detail for the two different models. Let us note already that a system satisfying only condition (i) will not necessarily exhibit SSB in the sense of Definition~\ref{def.ssb_textbook_finite}, the non-interacting spin-1/2 particles on a lattice, which have zero correlation between different sites, being a counterexample. Moreover, the requirement of long-range order ensures that symmetry-broken states satisfy already one of two stability requirements, namely that of dynamical stability under (time-dependent) perturbations of the Hamiltonian.

\paragraph{Discrete symmetry breaking}

For global discrete symmetries, Horsch and von der Linden have proven that a symmetric ground state with long range order is accompanied by a low-lying excited state \cite[Thm 2.2]{horschSpincorrelationsLowLying1988} whose excitation energy is smaller than $1/N$ times a constant and whose fluctuations of $\braket{O(x)}$ vanish in the thermodynamic limit. This means that \textit{in the thermodynamic limit} there exists a symmetry-broken state with ground state energy. Because of its low excitation energy $\Delta_N =\mathcal{O}(1/N)$, for finite system size the low-lying excited state can become the ground state when the Hamiltonian is perturbed by a perturbation of order $1/N$, which is extremely small for typical systems containing about $10^{23}$ particles. The perturbation then selects in which one of the symmetry-broken states the system is found.

As an example, consider again the transverse-field Ising model with $\mathbb{Z}_2$-symmetry generated by the operator in Eq.~\eqref{eq.z2_symmetry_ising}.
The quantum Ising model has two phases, driven by the transverse field strength $B$. For $B>1$ it has a paramagnetic ground state, where, in the limit of $B\to\infty$, the field pushes the spin at each site to the $\sigma_1$-eigenstate. For $B<1$ its ground state is a symmetric superposition $\ket{+_B}=(\ket{\uparrow_B}+\ket{\downarrow_B})\sqrt{2}$ of the dressed states
\begin{equation}
    \ket{\uparrow_B} := \frac{1}{\sqrt{1+NB^2}}\left(\ket{0\dots0} + B\sum_x\sigma_1(x)\ket{0\dots 0}\right),
\end{equation}
similarly defined for $\ket{\downarrow_B}$ \cite[]{pfeuty1970one, kogut1979introduction}. 
This state has long-range order since $$\braket{\sigma_3(x)\sigma_3(y)} = 1-\frac{4B^2}{1+NB^2},$$ for any $x\neq y$, which is independent of $x,y$. 
It will therefore be dynamically stable: if the system starts in this symmetric ground state, small symmetry-breaking perturbations introduced at later times will not drastically change the state. However, the symmetric state will not satisfy the cluster decomposition since a local measurement of $\sigma_3(x)$ will collapse the symmetric state on either $\ket{\uparrow_B}$ or $\ket{\downarrow_B}$. This means that even though the symmetric state $\ket{+_B}$ has long-range order, it is not expected to be found in any real-life scenario, where the environment can introduce such local measurements.

The question then becomes which states one does expect to find in real life. The answer can be found by looking at the energy behavior of the symmetry broken states.
The first excited state of $H_N$ (up to first order perturbation theory) is:
\begin{equation}
    \ket{-_B}:=(\ket{\uparrow_B}-\ket{\downarrow_B})/\sqrt{2}.
\end{equation}
By solving the quantum Ising model, one can show that the energy difference $\Delta_N$ between the ground state and the first excited state goes to zero with increasing system size
\cite[]{suzuki2012quantum}:
\begin{align}
    \Delta_N \approx (1-B^2)B^N.
\end{align}
This means that $\Delta_N \to 0$ exponentially in the thermodynamic limit\footnote{Notice that this scaling of the excitation energy is consistent with the upper bound $\Delta_N = \mathcal{O}(1/N)$ in \cite[]{horschSpincorrelationsLowLying1988}.} as long as the system is in a ferromagnetic phase with $0\leq B < 1$. Therefore, in this limit $\ket{\uparrow_B}$, $\ket{\downarrow_B}$, which are formed as a superposition of the ground and first excited state, have ground state energy, even though strictly speaking they are still no energy eigenstates. Moreover, contrary to $\ket{+_B}$, these states do satisfy the cluster decomposition and therefore are not only dynamically stable states, but also stable under local measurements.

The final question is then what causes the system to be in one of its symmetry-broken states.
As a consequence of the small energy-difference between the symmetric and symmetry-broken states, a local perturbation $-\epsilon \sigma_3(x)$ of the Hamiltonian $H_N$ of Eq. \eqref{eq.local_ising_hamiltonian} can push the system in one of the symmetry-broken states \cite[]{bogoliubov1961quasi}. Indeed, suppose that $N$ is finite and let $\epsilon > 0$ and of order $\Delta_N$. Then the ground state of $H_N$ is $\ket{\uparrow_B}$, which for $B<1$ has a nonzero average magnetization~\cite[]{pfeuty1970one},
\begin{equation}
    \braket{M} = \left\langle\frac{\sum_{x=1}^N \sigma_3(x)}{N}\right\rangle = \left(1-B^2\right)^{1/8}.
\end{equation}
Notably, if one first lets the system size increase to $N'>N$ and then the size of $\epsilon$ decrease to smaller $\Delta_{N'} < \Delta_{N}$, the system stays in a magnetized ground state.
Contrarily, if one assumes the system is in perfect isolation and there are no perturbations of any size $\epsilon$ present, the Hamiltonian is perfectly symmetric and has a symmetric ground state for any system size. Phrased differently, we find so-called \textit{singular} or \textit{non-commuting} limits of the global order parameter $\braket{M}$~\cite[]{bogoliubov1961quasi}:
\begin{equation}
    0 = \lim_{N\to\infty}\lim_{\epsilon \downarrow 0} \braket{M} \neq \lim_{\epsilon \downarrow 0}\lim_{N\to\infty} \braket{M} = (1-B^2)^{1/8}.
\end{equation}
These non-commuting limits illustrate the diverging susceptibility of the system's order parameter to symmetry-breaking perturbations in the thermodynamic limit, thus explaining how a large system can be found in a symmetry-broken state with nonzero order parameter. For this reason, non-commuting limits are often used in physics as a signature of SSB.

\paragraph{Continuous symmetry breaking}

When the global symmetry is continuous, a symmetric ground state with long range order has been shown to be accompanied by a so-called \textit{tower of states} \cite[]{tasakiLongrangeOrderTower2019}. This tower, first described in \cite{anderson1952approximate}, consists of states whose energy differences scale as $\propto 1/N$. Koma and Tasaki further argued why the states in the tower are physically natural: they satisfy the cluster decomposition and are thus stable in real-life situations \cite[]{koma1994symmetry,Tasaki2020}. In the thermodynamic limit of $N\to\infty$, the tower of states `collapses' in that the energy difference between states in the tower goes to zero. Then, the symmetry-broken states formed from this tower obtain ground state energy.\footnote{For the breaking of a continuous symmetry one furthermore needs the tower of states to be thinly occupied, such that its contribution to the free energy vanishes in the thermodynamic limit \cite[]{vanwezel2007spontaneous, vanwezelSpontaneousSymmetryBreaking2008}.
For this reason, the tower of states is also known as the \textit{thin spectrum}.} This means that the tiniest perturbation of the Hamiltonian will be enough to propel the system into a symmetry-broken ground state.

The BCS-model of superconductivity is an example of such a quantum system with a continuous global $U(1)$ symmetry. The reduced BCS Hamiltonian is defined as:
\begin{equation}
    H_{BCS} = \sum_{\mathbf{k}, \sigma} \epsilon_{\mathbf{k}} n_{\mathbf{k}, \sigma} - \frac{G}{V} \sum_{\mathbf{k}, \mathbf{l}} c_{\mathbf{k} \uparrow}^\dagger c_{-\mathbf{k} \downarrow}^\dagger c_{-\mathbf{l} \downarrow} c_{\mathbf{l} \uparrow},
\end{equation}
where $\epsilon_{\mathbf{k}}$ is the single-particle energy at momentum $\mathbf{k}$, $n_{\mathbf{k}, \sigma} = c_{\mathbf{k} \sigma}^\dagger c_{\mathbf{k} \sigma}$ is the number operator for electrons with spin $\sigma \in \{\pm 1\}$, $G > 0$ is the attractive pairing strength, $V$ is the system volume, and $c_{\mathbf{k} \sigma}^\dagger$ ($c_{\mathbf{k} \sigma}$) are the electron creation (annihilation) operators. This Hamiltonian commutes with the total particle number operator $\sum_{\mathbf{k}, \sigma} n_{\mathbf{k}, \sigma}$, which generates the global $U(1)$ gauge symmetry when exponentiated (more on this in Section~\ref{superconductor}). The order parameter of the superconducting phase is identified with the expectation value of the Cooper pair density, i.e. the density of electron pairs of opposite momentum and spin. 

By solving the model exactly, one can show that it possesses a tower of states of collective low-energy eigenstates \cite[]{vanwezelSpontaneousSymmetryBreaking2008}. The excitations correspond to the creation of Cooper pairs, which add vanishingly small energy when $N\to\infty$. The symmetry-broken states, formed by states in the tower, have nonzero expectation value of the order parameter. Because the tower of states collapses for large~$N$, the ground state of the system changes drastically when the BCS Hamiltonian is perturbed by a small symmetry-breaking field of the form $-\epsilon c_{\mathbf{k} \uparrow} c_{-\mathbf{k} \downarrow}$. In that case one again finds non-commuting limits as a signature of SSB \cite[]{bogoliubov1961quasi}:
\begin{equation}
    0 = \lim_{N\to\infty}\lim_{\epsilon \downarrow 0} \braket{c_{\mathbf{k} \uparrow} c_{-\mathbf{k}}} \neq \lim_{\epsilon \downarrow 0}\lim_{N\to\infty} \braket{c_{\mathbf{k} \uparrow} c_{-\mathbf{k}}} \neq 0.
\end{equation}

\subsection{Boundaries and perturbations}\label{sec:boundariesperturbations}

Thus far in our presentation of SSB we have not elaborately discussed boundary conditions. Here, we show that the role of boundaries is threefold: 
\begin{itemize}
    \item incorporating the fact that the system has an environment from which it can receive the perturbations necessary for SSB;
    \item encoding the requirement that the object ends and hence that the order parameter of the object goes to zero;
    \item allowing for an exterior reference frame which, by the relational interpretation of symmetries, is necessary for the physicality of the symmetry breaking in question.
\end{itemize}
The first point is a bit subtle. On the one hand, boundary conditions do serve to separate a system from its environment. Thus, they might seem to isolate the subsystem. However, boundary conditions can also encode influences from the environment --- influences which are always present for a realistic finite system. In classical systems, these influences are explicitly taken into account in the definition of a Gibbs measure \cite[]{Georgii}. However, in quantum systems they are not, and should be included by hand if one deems them relevant for the description of the system. Often, the role of the environment in the Hamiltonian of a system is implicit. For instance, the transverse field present in the quantum Ising model is usually not a feature of the ferromagnetic spin system itself, but represents a large external magnetic field which is present outside of the system. In the case of SSB, once one accepts that a perfectly symmetric Hamiltonian is an idealized description of a real-life system, embedded in an environment, the tiny $\mathcal{O}(1/N)$ perturbations required to `spontaneously' break the symmetry of finite systems can be naturally introduced by the environment. Here, a point of caution should be made. For magnets such as the transverse-field Ising model, it is very reasonable to expect an object, e.g. an oxygen molecule, with nonzero magnetic moment to be present when the magnet undergoes a phase transition, and thereby introduce the required symmetry-breaking perturbations. However, for BCS superconductors the symmetry-breaking field corresponds to a nonzero expectation value of Cooper pairs, which occur less naturally in the environment. In general, one needs a symmetry-breaking perturbation which \textit{couples to the order parameter of the system} in order to establish SSB~\cite[]{vanwezel2026ssb}. For a magnet such perturbations abound,
for a superconductor their origin is not as clear, and the dynamical emergence of a definite global \(U(1)\) phase remains less well-understood \cite[]{Greiter2005,vanwezelSpontaneousSymmetryBreaking2008}.\footnote{In our view, the position that a superconductor breaks its $U(1)$ symmetry only when another superconductor is brought in is contestable for two reasons. First of all, because of the long-range order of the symmetric state, the superconductor can only adiabatically move to a symmetry-broken state, thus resulting in a delayed Josephson current between the two superconductors. We are not aware of any experiments that have been done to verify this delay. Secondly, if the symmetry of the symmetric superconductor is broken by the other superconductor, one would (at least naively, by introducing the presence of a symmetry-broken superconductor as a perturbation of the form $-\epsilon c_{\mathbf{k} \uparrow} c_{-\mathbf{k} \downarrow}$ in the Hamiltonian of the symmetric superconductor) expect them to have the same $U(1)$ phase, thus resulting in no Josephson current at all. It would be interesting to investigate a less naive picture of the interaction between two superconductors and the resulting effect on the Josephson current.}

At the same time, for a finite system one has to deal with the fact that the system actually ends at a certain point. This means that outside of the system the order parameter must vanish. If the system is considered in isolation, the order parameter will vanish completely. For an ordinary magnet, for instance, the magnetic field, and therefore the order parameter, will decay as $1/r^3$. But realistically, the system is not perfectly isolated. This is the tension central to SSB: on the one hand, one studies the system as if it is isolated, requiring the order parameter to vanish on the boundary. On the other hand, one knows that, after removing the idealization of perfect isolation, the environment always induces perturbations on that boundary. Thus one must relax the strict requirement of the order parameter vanishing on the boundary. The boundary condition induced by the environment will then determine the stable symmetry-broken sector in which the system ends up.

This two-fold role of boundaries in SSB fits well with the relational view on symmetries developed by philosophers \cite[]{greavesEmpiricalConsequencesSymmetries2014,tehGalileoGaugeUnderstanding2016,gomesHolismEmpiricalSignificance2021,wallaceIsolatedSystemsTheireen,wallaceIsolatedSystemstwee,tehPhilosophySymmetry2024}. The relational interpretation states that symmetry transformations can have physical significance only \textit{relationally}, i.e. across a boundary separating subsystem from environment. The idea is often referred to by the \textit{Galileo's ship} thought experiment, in which one imagines a person in the cabin of a ship at sea. This person cannot tell whether the ship is boosted by  a constant velocity from within the cabin, but only in relation to the water or the shore. When the relational view is applied to symmetry-breaking systems, one has to conclude that one can tell apart the various directions in which a symmetry can be broken only in relation to other objects which are in a state that breaks the same symmetry. In Section~\ref{superconductor} we will see that this relational physical significance of symmetry breaking is in fact perfectly instantiated by superconductors.

\subsection{A note on spontaneity}

In this Section we have seen how for the Ising model and the BCS superconductor, as system size increases, the presence of a global symmetry and long range order make the symmetric ground state of a system increasingly unstable under small symmetry-breaking perturbations of its idealized, perfectly symmetric Hamiltonian.
This increased sensitivity explains why the idealized Hamiltonian gives incorrect predictions about the system's ground state, and why for real-life systems a symmetry-broken instead of a symmetric ground state is observed. In all cases, the system is required to be on its way to the thermodynamic limit. Otherwise, the two stability requirements are not met. When $N$ is small, the symmetry-broken states are not even approximately of ground state energy. Moreover, when $N$ is small, the system can tunnel to a different symmetry-broken state with probability $e^{-N} > 0$.

Thus, we have seen that the phenomenon called \textit{spontaneous symmetry breaking} is, despite its name, not truly spontaneous\footnote{An alternative name for SSB in finite systems has also been \textit{obscured symmetry breaking} \cite[]{koma1994symmetry}. In a way, this naming is enlightening, because the SSB of finite systems is simply not spontaneous, and the origin of the SSB in finite systems is \textit{obscured} by the idealized, perfectly symmetric Hamiltonian that represents the system.}~\cite[]{vanwezel2010broken,landsmanSpontaneousSymmetryBreaking2013,fraserSSB, Wallace:2018zbu, vanwezel2026ssb}, since, even in the thermodynamical limit some nonzero perturbation is necessary to introduce a preferred direction in which the system's state is broken. However, because the required symmetry-breaking perturbations are extremely small and decrease exponentially with system size, the mechanism is also inherently different from explicit symmetry breaking, where the symmetry-breaking `perturbations' are large fields that scale with system size.\footnote{For example, adding the transverse-field $B$ to the ferromagnetic Ising Hamiltonian (the first term in Eq. \eqref{eq.local_ising_hamiltonian}) explicitly breaks its global $\sigma_2$ and $\sigma_3$ spin-flip symmetry, but qualitatively affects the ground state only for $B$ larger than the ferromagnetic coupling and present at all lattice sites.} Therefore, we propose a new interpretation of the acronym SSB, namely \textit{subtle symmetry breaking}, which is itself only subtly different from the original nomenclature. For the remainder of this article we read `SSB' as subtle symmetry breaking, which is only seemingly, but not truly spontaneous.\footnote{We thank Klaas Landsman for suggesting the terminology `subtle' symmetry breaking.}
\section{The Thermodynamic Idealization}\label{secthermoidealization}

In the previous Section our treatment of SSB was expressly `physical' and focused on the underlying mechanism for systems of finite size. If one wants to push the explanation to systems of infinite size, for example to describe SSB on an infinitely large spacetime, then the Hilbert space of states and the algebra of observables become mathematically ill-defined. To remedy this, a sophisticated, algebraic tradition of mathematical SSB has been developed \cite[]{landsmanFoundationsQuantumTheory2017}. Here, we show how algebraic definitions of SSB, which apply to quantum systems with an infinite number of degrees of freedom only, use the robustness and low energy of symmetry-broken states to define SSB based on superselection sectors. These sectors are encoded naturally in completeness conditions on the algebra of observables, which result in a C$^*$-algebra of (quasi-)local observables that makes it impossible to go from one sector to another via operators in the algebra.
Furthermore, we show how asymptotic boundary conditions on the states effectively isolate the system, thereby hiding the perturbative boundary mechanism that explains why, in nature, a finite system is always found in one of its symmetry-broken sectors.

\subsection{Algebraic quantum theory}\label{sec:algebraic}

In algebraic approaches to quantum theory \cite[]{haagLocalQuantumPhysics1996,bratteli1998operator,strocchiSymmetryBreaking2008, ruetscheInterpretingQuantumTheories2011, landsmanFoundationsQuantumTheory2017} the aim is to formalize quantum theory and in particular quantum systems with infinitely many degrees of freedom. This formalization has led to multiple definitions of SSB. Here, we briefly introduce the terminology needed to formulate these definitions.

In the algebraic formalism, observables are the self-adjoint elements (i.e. those satisfying $a^*=a$) of an abstract C$^*$-algebra $\mathcal{A}$ instead of simply $B(\mathcal{H})$, the bounded operators on the state space $\mathcal{H}$. For a lattice spin-1/2 system, this would be the C$^*$-algebra generated by all local spin operators, i.e. Pauli matrices. Here \textit{local} means acting on finitely many lattice sites. 
The resulting C$^*$-algebra, because of completeness conditions, contains limit points in addition to all local operators, and is called the \textit{quasi-local algebra of observables}.\footnote{For details on how to construct this algebra, see e.g. \cite[]{landsmanClassicalQuantum2005} or \cite[]{strocchiSymmetryBreaking2008}.} This is similar to how the algebra $C_c(X)$ of compactly supported functions on a locally compact space $X$ is completed to $C_0(X)$, the algebra of continuous functions.
A \textit{state} in the C$^*$-algebraic formalism is a linear map $\omega\colon \mathcal{A}\to\mathbb{C}$, i.e. an assignment of expectation values to the operators. For this map to be an algebraic state it must be positive: $\omega(a^*a)\geq 0$ for all $a\neq 0$, and have norm one. In the algebraic formalism, a state $\omega$ is called \textit{mixed}\footnote{This is a somewhat misleading definition as it suggests that such states are also mixed in the quantum mechanical use of `mixed state', i.e. as if they represent either an ensemble of pure states or a subsystem that is entangled with a larger system. In the algebraic definition, this is however not the case. The algebraic state induced by the quantum mechanically pure state ``all up + all down'' is mixed in the algebraic sense, since no observables in the algebra connect the ``all up'' to the ``all down'' state.} if it is a nontrivial convex sum of states, i.e. one can write it as
\begin{equation}
    \omega(A) = \lambda \omega_1(A) + (1-\lambda)\omega_2(A)
\end{equation}
for some $\lambda \in (0,1) \subseteq \mathbb{R}$, with $\omega_1\neq\omega_2$.
A \textit{symmetry} in algebraic quantum theory is a $*$-isomorphism $\gamma\colon\mathcal{A}\to\mathcal{A}$, i.e. an automorphism of the algebra that preserves all the relevant structure, including the linear structure, the algebra structure, and the involution.\footnote{These latter two statements just mean that $\gamma(ab)=\gamma(a)\gamma(b)$ and $\gamma(a^*)=\gamma(a)^*$ for all $a,b\in\mathcal{A}$.} 
The \textit{dynamics} of a quantum system with C$^*$-algebra $\mathcal{A}$ is determined from the local Hamiltonian $H_N$ of the finite system, which was originally defined on $N^d$ lattice sites. Indeed, when $H_N$ is local (as defined in Section~\ref{sec.textbookdefn}), for each $a \in \mathcal{A}, t \in \mathbb{R}$ it induces a $*$-automorphism on $\mathcal{A}$ given by \cite[Thm 9.15]{landsmanFoundationsQuantumTheory2017}
\begin{equation}\label{eq.dynamics}
    \alpha_t(a) = \lim_{N\to\infty} e^{it H_N} a e^{-itH_N}.
\end{equation}
The resulting $1$-parameter family of $*$-automorphisms $\alpha\colon \mathbb{R}\to \text{Aut}(\mathcal{A}), t \mapsto \alpha_t$, gives the system's dynamics in the Heisenberg picture.
Finally, we say that a system with dynamics $\alpha: \mathbb{R} \to \text{Aut}(\mathcal{A})$ has a \textit{symmetry group} $G$, if for any $g \in G$ and for all $t \in \mathbb{R}$
\begin{equation}
    \alpha_t \circ \gamma_g = \gamma_g \circ \alpha_t,
\end{equation}
in other words if $\gamma_g$ commutes with the dynamics for all $g$.

Within this algebraic framework, the stability of a state $\omega$ is defined by its behavior under the dynamics $\alpha_t$, leading to the formal identification of equilibrium states. For a system at zero temperature, a state is a ground state if it satisfies the energy condition $\omega(a^* \delta(a)) \geq 0$ for all $a$ in the domain of the derivation $\delta$, where $\delta$ is the generator of the dynamics $\alpha_t = e^{it\delta}$ (intuitively, this requirement states that any alteration to the system induced by the operator $a$ can only change the energy positively). At finite temperature $\beta = (k_B T)^{-1}$, stable states are characterized as KMS (Kubo-Martin-Schwinger) states , which satisfy the condition $\omega(a \alpha_{i\beta}(b)) = \omega(ba)$ for all $a, b \in \mathcal{A}$ \cite[]{kubo1957statistica,martin1959j,haag1967equilibrium,bratteli1998operator}. Unlike the physical notion of stability under local measurements and perturbations used in finite systems, these algebraic conditions ensure that the state is a global extremum of the relevant thermodynamic potential and is robust against perturbations of the dynamics in the thermodynamic limit, i.e. is what we previously called `dynamically stable'. 

The structure of the algebra $\mathcal{A}$ provides the basis for \textit{superselection}\footnote{For a detailed study of various definitions of superselection sectors see \cite{earmanSuperselectionRulesPhilosophers2008}.}: transitions between distinct sectors cannot be induced by any (quasi-)local observable $a \in \mathcal{A}$.\footnote{It is sometimes said that ``infinite energy'' would be required, which is correct in the sense that a nonzero field has to be applied at infinitely many lattice sites and even then, moving from one sector to another would happen adiabatically and thus take an infinite amount of time.} Physically, this implies that a state is stable under local measurements, as its expectation values remain confined to a single sector. This stability is captured mathematically by the unitary inequivalence of \textit{GNS representations}, which is a way to represent the C$^*$-algebra $\mathcal{A}$ on a Hilbert space $\mathcal{H}$, i.e. to revert back to standard quantum mechanics. The GNS construction takes a specific state $\omega\colon \mathcal{A}\to\mathbb{C}$ and produces a representation\footnote{Here $B(\mathcal{H})$ denotes the algebra of bounded operators on $\mathcal{H}$.}  $\pi_\omega\colon\mathcal{A}\to B(\mathcal{H}_\omega)$ on a Hilbert space $\mathcal{H}_\omega$ that is ``built up'' from the original state $\omega$, which itself is represented as a ``vacuum state'' $\ket{\Omega}\in \mathcal{H}_\omega$. The Hilbert space $\mathcal{H}_\omega$ consists of all states that differ only (quasi-)locally from the original state $\omega$. Two states $\omega$ and $\omega'$ belong to different superselection sectors if and only if there exists no unitary operator $U: \mathcal{H}_\omega \to \mathcal{H}_{\omega'}$ such that $U \pi_\omega(a) = \pi_{\omega'}(a) U$ for all $a \in \mathcal{A}$. Consequently, if a symmetry transformation $\gamma_g$ acts globally, altering the state at infinitely many sites, the transformed state $\omega \circ \gamma_g$ will yield a GNS representation unitarily inequivalent to that of $\omega$. This inequivalence is the basis for formalizing SSB in the thermodynamic limit: the symmetry is ``broken'' when the state space is partitioned into disjoint Hilbert space sectors, preventing any physical process described by $\mathcal{A}$ from restoring the symmetry or forming a coherent superposition between different symmetry-broken sectors. More precisely, this means that two states lying in different sectors automatically produce a mixed state when superposed.

In the framework of Doplicher-Haag-Roberts (DHR) superselection in algebraic QFT \cite[]{doplicherFieldsObservablesGauge1969,doplicherFieldsObservablesGauge1969a,doplicherLocalObservablesParticle1971,doplicherNewDualityTheory1989}, superselection sectors represent all states with the same conserved quantum numbers associated to some symmetry group action, think e.g. of charge or any symmetry-induced conserved quantity. States in different DHR sectors are non-superposable because they carry different conserved quantum numbers. This means that the symmetry transformation responsible for those conserved quantities acts differently on states in different subspaces, prohibiting a pure superposition of such states. To derive the results of DHR theory, one assumes the so-called \textit{selection criterion}, which states that all expectation values (of all observables) should approach the vacuum expectation values uniformly when the region of measurement is moved away from the origin. Clearly, the superselection criterion is an implementation of asymptotic boundary conditions, encoded in the fall-off of expectation values as compared to the vacuum expectation value. Thus, this provides a concrete example of how superselection arises from asymptotic boundary conditions in quantum theory. 

Similarly, for systems with SSB, different symmetry-broken states become non-superposable in the thermodynamic limit. This deviates from the finite-size description of SSB, where two different symmetry-broken states live in the same Hilbert space and can always be superposed. Here ``non-superposability'' of symmetry-broken states refers to the fact that such superpositions are mixed states and are therefore unstable against local measurements and do not satisfy cluster decomposition. In this way, this physical requirement for stability is implicitly incorporated in the algebraic superselection structure.

\subsection{Algebraic definitions of SSB}

The algebraic formalization has led to multiple definitions of SSB, of which we discuss three.
We view the definitions listed here as attempts at \textit{essential definitions}, i.e. definitions that are meant to capture a concept (SSB) which is preformal \cite[]{landsmanrigour}, as evinced by the fact that SSB is observed experimentally and used widely in a heuristic fashion in theoretical physics. Consequently, these definitions and any others that try to formalize an empirical concept should be viewed merely as \textit{hypothetical}.
That is, we may use them as a basis of rigorous, mathematical reasoning. However, the resulting mathematical statements, e.g. about where and when SSB may occur, can only be \textit{compared} to experimental reality and no conclusions about reality itself can be drawn from them. 
This also means that, ultimately, there is no guarantee that any definition of SSB will fully capture the phenomenon.
We will see that the algebraic definitions rigorously describe the phenomenon for infinite systems but are applicable to hardly any finite system. If SSB is understood as a phenomenon occurring for systems of finite size too, these definitions evidently suffer from an idealization problem.

In the thermodynamic limit of usual quantum systems, neither the Hamiltonian nor the state space $\mathcal{H}$ are well-defined. This makes Definition~\ref{def.ssb_textbook_finite} unsuitable as a rigorous definition of SSB for infinite quantum systems. To remedy this, we straightforwardly reformulate Definition~\ref{def.ssb_textbook_finite} in the $C^*$-algebraic approach to quantum theory.
\\
\begin{definition}\label{def.ssb_textbook_infinite}
    Consider a system with time evolution $\alpha: \mathbb{R} \to \text{Aut}(\mathcal{A})$ on the C$^*$-algebra $\mathcal{A}$. Let $G$ be a symmetry-group of the system. Let $\omega: \mathcal{A}\to \mathbb{C}$ be a stable state on $\mathcal{A}$. Then $G$ is said to be spontaneously broken by $\omega$ if
    \begin{equation}
        \exists g\in G: \omega \circ \gamma_g \neq \omega.
    \end{equation}
\end{definition}
\noindent This definition uses the fact that a stable state is mapped to another stable state by a symmetry transformation. However, it does not apply to general finite systems, for the same reason that Definition~\ref{def.ssb_textbook_finite} does not apply, namely that the symmetry-broken state has to be stable, which in the algebraic framework means that it is a ground- or KMS-state. As we have mentioned previously, for general finite systems, the symmetry-broken states are no exact ground states or even energy eigenstates (the transverse-field Ising model with zero field-strength being a notorious exception). Thus most of these systems do not satisfy the requirements of Definition~\ref{def.ssb_textbook_infinite}.

There exists a second definition of SSB in the algebraic formalism which is commonly used by mathematical physicists and philosophers,
centered around the fact that in the thermodynamic limit of a quantum system, no local operator can transform between distinct asymmetric ground states, the reason being that these differ by a \textit{global} operation, which for an infinite system acts non-trivially on infinitely many sites. 
\\
\begin{definition}\label{def.ssb_algebraic_GNS}
    Consider a system with time evolution $\alpha:\mathbb{R}\to\text{Aut}(\mathcal{A})$ and symmetry-group $G$, inducing $\gamma_g \in Aut(\mathcal{A})$ for all $g \in G$.
    The symmetry is said to be spontaneously broken by a stable state \text{$\omega\colon\mathcal{A}\to\mathbb{C}$} if there exists a $g \in G$ such that \text{$\gamma_g\colon\mathcal{A}\to\mathcal{A}$} is not unitarily implementable in the GNS representation \text{$\pi_\omega\colon\mathcal{A}\to B(\mathcal{H}_\omega)$}, i.e. there does not exist a unitary map $U:\mathcal{H}_\omega\to\mathcal{H}_\omega$ satisfying
    \begin{align*}
        \pi_\omega(\gamma_g(a))=U^*\pi_\omega(a)U,\;\;\;\;\;\text{for all }a\in\mathcal{A}.
    \end{align*}
    Equivalently, we may say that $\omega$ spontaneously breaks the symmetry $\gamma_g$ if $\omega$ and $\omega\circ\gamma_g$ generate unitarily inequivalent GNS representations \cite[]{strocchiSymmetryBreaking2008,landsmanFoundationsQuantumTheory2017}.
\end{definition}
\hfill \break
\noindent This definition moves from the initial idea that a symmetry $\gamma_g$ is spontaneously broken if $\omega\neq \omega\circ\gamma_g$, to the much stronger and more involved notion of a state and its symmetry-transformed counterparts living in different superselection sectors, related by a global symmetry transformation.  
Definition~\ref{def.ssb_algebraic_GNS} applies only to systems whose algebra allows unitarily inequivalent representations in the first place, i.e. systems that are of infinite size or have a non-simple\footnote{A C$^*$-algebra is called \textit{simple} if it has no two-sided, closed, nontrivial ideals.} C$^*$-algebra of observables. The Stone-von Neumann theorem guarantees that the usual finite quantum systems on $\mathbb{R}^n$ have a unique representation up to unitary equivalence. Thus, for finite systems with a simple algebra of observables, any symmetry is unitarily implementable in the GNS representation of any state $\omega\colon\mathcal{A}\to\mathbb{C}$ and SSB never occurs according to Definition~\ref{def.ssb_algebraic_GNS}.

The last algebraic definition of SSB that we discuss and, for mathematical simplicity, give at zero temperature only,\footnote{For nonzero temperature, the definition defines SSB to occur when there exist no \textit{primary} KMS-states.} is as follows.
\\
\begin{definition}\label{def.ssb_algebraic_pure}
    A system with time-evolution $\alpha_t: \mathbb{R}\to\text{Aut}(\mathcal{A})$ on a C$^*$-algebra $\mathcal{A}$ and symmetry group $G$ (i.e. $\alpha_t$ commutes with the $*$-automorphism $\gamma_g \in \text{Aut}(\mathcal{A})$ implementing the symmetry for any $g\in G$) is said to spontaneously break the symmetry $G$ at zero temperature if there exist no pure $G$-invariant (i.e. symmetric) ground states \cite[]{landsmanFoundationsQuantumTheory2017}.
\end{definition}
\hfill \break
\noindent To understand this definition, we need to examine the somewhat subtle requirement that there exist no \textit{pure} $G$-invariant ground states. 
Recall that a state $\omega$ is mixed if it is a nontrivial convex sum of states. For any infinite system exhibiting SSB, the symmetry-broken states are macroscopically distinct, and thus lie in different superselection sectors. Since the (typically unique) symmetric ground state is formed as a non-trivial linear combination of symmetry-broken states, but no observable connects the symmetry-broken states, the symmetric ground state will be \textit{mixed} (i.e. not pure) in the algebraic sense. Consequently, according to Definition \ref{def.ssb_algebraic_pure}, no pure symmetric ground states exist. A beautiful result in algebraic quantum theory states that precisely the algebraically pure (or primary) states satisfy the cluster decomposition~\cite[Cor 8.22 \& Thm 8.23]{landsmanFoundationsQuantumTheory2017}, meaning that they are stable under local measurements. One can then rephrase Definition~\ref{def.ssb_algebraic_pure} as: \textit{a system spontaneously breaks the symmetry of its dynamics if its symmetric ground state is not pure (or primary), and therefore non-clustering and unstable.}

Thus, we see how SSB for infinitely large systems can be rigorously defined in various ways. The algebraic Definitions~\ref{def.ssb_algebraic_GNS}-\ref{def.ssb_algebraic_pure} elegantly use the structure of the quasi-local algebra of observables to identify superselection sectors of states that are both dynamically stable and clustering. Because of their stability and ground state energy, these states will be the physically natural ground states of the infinite system. However, because the definitions require a nontrivial superselection structure, they only apply to infinite systems, which do not satisfy Earman's principle\footnote{This principle states that ``While idealizations are useful and, perhaps, even essential to progress in
physics, a sound principle of interpretation would seem to be that no effect
can be counted as a genuine physical effect if it disappears when the idealizations are removed'' \cite[]{earmanCuriePrincipleSpontaneous2004}.}: as soon as the idealization of working in the thermodynamic limit is removed, the definitions of SSB no longer apply \cite[]{landsmanSpontaneousSymmetryBreaking2013}. 

So far, however, the thermodynamic idealization has been understood solely in terms of the dimensionality of the quantum system in question. We will now show that the thermodynamic limit used in Definition \ref{def.ssb_algebraic_GNS} can alternatively be thought of as the idealization of locating the boundary of the system ``at infinity'', i.e. by implementing asymptotic boundary conditions.

\subsection{The role of asymptotic boundary conditions}\label{sec:roleasymptoticBC}

In Section~\ref{sec:boundariesperturbations}, we identified three primary roles of boundary conditions in finite systems:
\begin{itemize}
    \item incorporating the environment as a source of perturbations;
    \item demarcating a subsystem beyond which the order parameter vanishes; 
    \item enabling the relational physicality of symmetries and by extension symmetry breaking.
\end{itemize}
For infinite systems in the algebraic formalism, the first role disappears since the environment is located ``at infinity''. However, the latter two roles are preserved through asymptotic boundary conditions, which, as we will now explain, effectively isolate the system from the environment at infinity, and result in superselection sectors of stable, symmetry-broken ground states which differ only relationally, in that there is no absolute preferred sector.

The implementation of asymptotic boundary conditions is implicit in the GNS representation of a quasi-local algebra. By constructing a Hilbert space $\mathcal{H}_\omega$ from all states arising via finitely many excitations of some ``vacuum" state $\omega$, one restricts the physical state space to states that asymptotically approach the vacuum configuration at sufficient rapidity, at least on the dense subspace generated by quasi-local observables. Anticipating the relation between finite energy and asymptotic conditions in field theories in the next Section, one can roughly think of this as a condition on the energy of GNS states: if the Hamiltonian is normalized to assign zero energy to the GNS vacuum, the GNS states that differ only locally from this vacuum carry finite energy. This does not mean that \textit{every} vector in $\mathcal{H}_\omega$ has finite energy above the vacuum: firstly, there are GNS states corresponding to a quasi-local element of the algebra which is not strictly local, and secondly the Hilbert-space closure may contain limits that fail to lie in the domain of the Hamiltonian or have infinite energy expectation (after all, the Hamiltonian is an unbounded operator). Nonetheless, the GNS representation determines which states asymptotically approach the corresponding GNS vacuum sufficiently quickly in order to be allowed as members of that representation. This procedure effectively decouples the system from any environment at infinity, since local or quasi-local excitations of the vacuum do not extend out far enough towards infinity to count as global operations.

Crucially, these asymptotic conditions in the GNS representation do not need to be imposed by hand; they are intrinsic to the structure of the quasi-local algebra $\mathcal{A}$. Constructed as the inductive limit of finite-volume algebras $\mathcal{A}_N$, the algebra $\mathcal{A}$ contains operators acting on any finite number of sites, and, in its norm closure, only those observables that can be approximated by such local operators. Consequently, global operations---such as the $\mathbb{Z}_2$ spin-flip in the Ising model that acts on all sites simultaneously---exist only as automorphisms of $\mathcal{A}$ and not as elements within it (i.e., they are not inner automorphisms), leading to symmetry-related superselection sectors. The quasi-local structure also ensures that the dynamics $\alpha_t \in \text{Aut}(\mathcal{A})$ defined by Eq.~\eqref{eq.dynamics} are well defined in the $N\to\infty$ limit (even if the Hamiltonian $H_N$ were not normalized, as the constant energy shift drops out of the commutator with any local observable, and hence does not affect $\alpha_t$). 

For a concrete example, consider the transverse-field Ising model from Section~\ref{sec.defining_ssb}. Let us assign zero energy to the all-up or all-down ground states. In comparison, a state with alternating spins in the $N\to\infty$ limit carries infinite energy because a unit of energy is associated with every misaligned pair \textit{ad infinitum}. Because of the structure of the quasi-local algebra, such an infinitely alternating state cannot be part of the GNS representation of the all-up or all-down ground state. The elements from which the GNS representation is constructed have approximately finite tails for $N\to\infty$: as one moves toward infinity, the spins must eventually align with the chosen vacuum. The Hilbert space consists not of $\ell^2$ over all spin configurations, but only of a restricted space of configurations, together with its Hilbert-space closure. Thus the infinitely alternating state and the all-up state can never be part of the same GNS representation, and the root of this fact is that these states differ at infinitely many sites.

In conclusion, the asymptotic fall-off inherent in the quasi-local algebra gives rise to distinct superselection sectors generated by symmetry-broken states of ground state energy. The quasilocal algebra does not in itself specify \textit{which} state is taken as the vacuum state, from which other states can only differ quasi-locally. This choice of vacuum also cannot be induced by a perturbation from the environment because the environment has been pushed out to infinity, making the system effectively isolated. Consequently, the symmetry-broken sector in which the system is found, has to be imposed by hand and the symmetry seems to be broken spontaneously. 
\section{SSB in field theories}\label{relationalsymmetries}

In the previous Sections we explained that boundaries are central to SSB in both finite and infinite systems and used two lattice models as examples. In this Section we explain that boundaries play very similar roles for field theories. In fact, the idealization problem that arises when moving from finite to infinite lattices, and thereby from finite to asymptotic boundaries, also appears in field theories, both classical and quantum ones.

To argue this point we introduce the classical Abelian Higgs model on both infinite Minkowski spacetime (Section~\ref{sec:minkowski}) and on a finite region (Section~\ref{superconductor}). We thereby find some kind of ``classical superselection'' structure arising inevitably from the boundaries.

\subsection{The Abelian Higgs model in Minkowski spacetime}\label{sec:minkowski}

The Abelian Higgs model \cite[]{higgsBrokenSymmetriesMasses1964,Higgs:1964ia,englertBrokenSymmetryMass1964,guralnikGlobalConservationLaws1964} consists of an electromagnetic field coupled to a scalar Higgs field. Its quantum-field theoretical foundations on Minkowski spacetime are relatively well-understood \cite[]{strocchiIntroductionNonPerturbativeFoundations2016}, though not at the level of rigor desired in AQFT. Also, it is contested whether the universe really experienced some sort of dynamical electroweak phase transition shortly after the Big Bang.

The Lagrangian of the Abelian Higgs model on Minkowski spacetime is given by
\begin{align*}
    \mathcal{L}=-\frac{1}{4}F^{\mu\nu}F_{\mu\nu}-(D_\mu)^*\varphi D^\mu\varphi-V(\varphi),
\end{align*}
where $F_{\mu\nu}$ is the electromagnetic field strength tensor of the gauge field $A_\mu$, $D_\mu$ the covariant derivative and $V(\varphi)$ the (Mexican hat) potential of the Higgs field. Crucially, the Lagrangian exhibits a gauge symmetry given by $A_\mu\to A_\mu+\partial_\mu\lambda, \varphi\to e^{i\lambda}\varphi$. The well-known standard narrative of the Higgs mechanism then starts from an expansion around a minimum $\varphi_0=v/\sqrt{2}$ of $V(\varphi)$ and derives a new expression of $\mathcal{L}$ that contains a massive gauge boson. The choice of minimum ``breaks'' the $U(1)$ gauge symmetry of the theory, which is why it is said that the Higgs mechanism functions through gauge symmetry breaking.

This standard narrative has been contested at length \cite[]{earmanLawsSymmetrySymmetry2004,smeenkElusiveHiggsMechanism2006,lyreDoesHiggsMechanism2008,struyveGaugeInvariantAccounts2011a,friederichGaugeSymmetryBreaking2013,berghoferGaugeSymmetriesSymmetry2023}, leading to the more refined understanding that the physical part of gauge symmetry breaking is the breaking of the \textit{global} $U(1)$ symmetry, corresponding to constant gauge parameters~$\lambda$ \cite[]{borsboomdeharo}.

Crucially, the physical difference between global and local gauge symmetries only arises because of the presence of asymptotic boundary conditions on the fields $F_{\mu\nu}, A_\mu$ and $\varphi$ \cite[]{borsboomdeharo}. Indeed, on a Cauchy surface in Minkowski spacetime one needs to impose fall-off conditions on electric, magnetic, gauge and Higgs fields in order to ensure a well-defined Lagrangian, Hamiltonian and Legendre transform between them. These boundary conditions must be respected by the gauge group acting on the fields, leading in turn to asymptotic conditions on the gauge transformations themselves.\footnote{For details see e.g. \cite[]{SNIATYCKI1988291,sniatycskiboundarycondspatboundeddom,binz,Giulini:1994bi,lusannaDiracObservablesHiggs1997,struyveGaugeInvariantAccounts2011a,tehGalileoGaugeUnderstanding2016,gomesUnifiedGeometricFramework2018,gomesUnifiedGeometricFramework2019,gomesGaugingBoundaryFieldspace2019,gomesQuasilocalDegreesFreedom2021,rielloEdgeModesEdge2021,wallaceIsolatedSystemstwee,rielloHamiltonianGaugeTheory2024,rielloNullHamiltonianYang2025,Borsboom:2025agn,borsboomdeharo}.} Loosely speaking, since the electric, magnetic and gauge fields are required to vanish at infinity, gauge transformations must approach a constant towards spatial infinity at a fast enough rate, to ensure that their derivative is zero so that $A_\mu$ remains unchanged. However, in the theory of constrained Hamiltonian systems\footnote{See \cite[]{binz,henneauxQuantizationGaugeSystems1992,marsdenIntroductionMechanicsSymmetry1999,Gotay:1997eg,Gotay:2004ib,Gotay2006MomentumMA}} with boundary conditions, only gauge transformations that become the identity at spatial infinity are generated by the constraints and are interpreted as redundant. Thus there is a residual asymptotic global gauge group isomorphic to $U(1)$ which is physical, namely the quotient of the group $\mathcal{G}^I$ of asymptotically constant gauge transformations (which leave the boundary conditions \textit{Invariant}) by the group $\mathcal{G}^\infty$ of asymptotically trivial gauge transformations vanishing at infinity (which are interpreted as unphysical).\footnote{This interplay between allowed and trivial gauge transformations is very popular in current high energy physics, where so-called \textit{asymptotic symmetry groups} are studied in the context of holography. See e.g. \cite[]{stromingerAsymptoticSymmetriesYangMills2014,stromingerLecturesInfraredStructure2018,henneauxAsymptoticSymmetriesElectromagnetism2018,pasterskiCelestialHolography2021,Tanzi:2020fmt,rielloNullHamiltonianYang2025}. By finding e.g. the asymptotic symmetry group of the gravitational field in asymptotically flat spacetime, one may hope to construct a dual quantum field theory on the conformal boundary whose symmetry group is that asymptotic symmetry group. The most famous example is the BMS group on a null boundary \cite[]{bondiGravitationalWavesGeneral1962,sachsAsymptoticSymmetriesGravitational1962}.} Symmetry breaking in the Abelian Higgs model on Minkowski spacetime can then be understood as a breaking of this global gauge group $\mathcal{G}^I/\mathcal{G}^\infty\cong U(1)$. Physical significance here is understood in a relational sense, where the subsystem-environment distinction is between space (represented by the Cauchy surface) and spatial infinity (the conformal boundary of the Cauchy surface).

Importantly, the global $U(1)$ gauge symmetries can only be detected when a charged matter field (such as the Higgs field) is present, since such a field actually transforms (with a constant phase) under a global gauge transformation. The gauge field $A_\mu$ itself does not change because of the derivative $\partial_\mu\lambda$. In other words: the Higgs field is an order parameter for the \textit{global} $U(1)$ gauge symmetry breaking. One has a choice as to what one calls the `vacuum': it could either be a zero value of the Higgs field, or a minimum of the Mexican hat potential. The zero value is unstable, whereas the minima are stable under small fluctuations. This is a classical analog of what happens in the Ising model, where one can build up the GNS representation either from the mixed, non-clustering and hence unstable superposition $\ket{+}=(\ket{\uparrow}+\ket{\downarrow})/\sqrt{2}$, or from the stable $\ket{\uparrow}$ or $\ket{\downarrow}$ states. 

So far this is a purely classical story, but it can be extended to axiomatic QFT by means of the Wightman approach \cite[]{morchioLocalizationSymmetries2007,strocchiIntroductionNonPerturbativeFoundations2016,strocchiSymmetryBreakingStandard2019,borsboom2024spontaneous,borsboomdeharo}. In the quantization of the Abelian Higgs model, two types of superselection structure appear. The first consists of the Doplicher-Haag-Roberts \cite[]{doplicherFieldsObservablesGauge1969,doplicherFieldsObservablesGauge1969a,doplicherWhyThereField1990} charge sectors labelled (for $U(1)$) by elements of $\mathbb{Z}$. These charge sectors exist only in the unbroken phase, when the global $U(1)$ gauge symmetry is still intact \cite[]{morchioLocalizationSymmetries2007,strocchiIntroductionNonPerturbativeFoundations2016}. They also have a classical analogue \cite[]{rielloHamiltonianGaugeTheory2024}, known as \textit{flux superselection sectors}. The second type are the sectors corresponding to directions of global $U(1)$ symmetry breaking in the sense of Definition~\ref{def.ssb_algebraic_GNS}. In other words: the superselection sectors correspond to the different directions in which the global $U(1)$ gauge symmetry can be broken, which is why they are relevant for our study of SSB.

But, similarly to the quasi-local algebra in Section \ref{sec:roleasymptoticBC}, these superselection sectors exist only because of the presence of boundary conditions, viz. the fall-off conditions at spatial infinity. If one imposes the asymptotic boundary condition in which the Higgs field approaches a covariantly constant minimum, then the global $U(1)$ symmetry is broken and one is forced into a particular $U(1)$ ``superselection'' sector. Heuristically speaking, a ``global operation'' is required to go from sector to sector, since one would need to change the Higgs field all the way at spatial infinity to do so. The underlying idealization is therefore the asymptotic (and thus possibly unrealistic) nature of the boundary conditions, rather than the mere fact that the quantum system in question is finite- or infinite-dimensional. Indeed, a quantum field already contains an infinite number of degrees of freedom in a bounded space, so the idealization that leads to a superselection structure here is clearly not (only) the one of moving to an infinite-dimensional Hilbert space, but rather that of extending the fields out towards infinity while still requiring finiteness of energy.

\subsection{Superconductors and the Josephson effect}\label{superconductor}

Let us therefore consider what happens when one does not make the thermodynamic idealization. Superconductors provide a useful case study of such a finite-size application of the Abelian Higgs model. In fact, superconductors predated the Higgs mechanism in particle physics and were actually its inspiration \cite[]{nambuDynamicalModelElementary1961,nambuDynamicalModelElementary1961a}, though the analogy has been argued against because, unlike in superconductors, the Higgs mechanism in particle physics is not well-understood as a dynamical mechanism with accompanying phase transition~\cite[]{fraserHiggsMechanismSuperconductivity2016}. On the other hand, physicists do look for empirical signatures of such a phase transition, for instance in the guise of domain formation and cosmic strings \cite[]{Kibble:1976sj,Vilenkin:2000jqa}.

For superconductors the Ginzburg-Landau (GL) \cite[]{Ginzburg:1950sr} order parameter $\psi$ functions like the Higgs field (derived microscopically as the density of Cooper pairs) in that it is coupled to the electromagnetic field through a covariant derivative. Its nonzero vacuum value breaks the global $U(1)$ gauge symmetry, generated by the BCS number operator defined in Section~\ref{ssec.mechanism}, and gives photons inside the superconductor an effective mass. 

Despite their finite extent superconductors provide empirical evidence for the idea that the global $U(1)$  gauge symmetry group of the Abelian Higgs model is physical, namely via the \textit{Josephson current} \cite[]{josephsonPossibleNewEffects1962}. This current flows between two superconductors that are brought close together, and depends only on their \textit{relative} global $U(1)$ phase difference. Thus, although it does not allow one to observe the \textit{absolute} phase of a Ginzburg-Landau order parameter, it clearly exhibits the relational character of the physical $U(1)$ symmetry. Crucially, the Josephson current flows across the \textit{boundary} between two superconductors. This underscores that it is indeed the presence of a boundary that makes the global gauge group of the Abelian Higgs model physical, even in the case of a finite system. In the language of the \textit{Galileo's ship} thought experiment, one superconductor plays the role of the ship's cabin and the other of the reference sea/shore. The GL phase is analogous to the velocity of the ship. From ``inside'' the superconductor one does not notice in which direction the GL phase points, but from an environment separated by a boundary one does notice a difference in this phase. When the boundary between two superconductors is eliminated and the phases align, they form one big superconductor, and there is no Josephson current.

However, the nature of the boundary of a finite superconductor is different from the asymptotic boundary conditions of the Higgs model on Minkowski spacetime. For a physical superconductor the electric and magnetic fields do \textit{not} vanish on the boundary, but only inside the superconductor (for the magnetic field this is due to the Meissner effect), whereas on Minkowski spacetime electric and magnetic fields were required to fall-off asymptotically to ensure finite energy. Furthermore, because the physical superconductor has an actual end, the order parameter does need to vanish outside this boundary, since there one can no longer support a Ginzburg-Landau phase, unless a second superconductor is brought in. The question is what happens when two superconductors, both with their own GL order parameters, are first separated by a boundary on which the order parameter vanishes, and subsequently are brought together. This situation is known as the problem of \textit{gluing} \cite[]{gomesHolismEmpiricalSignificance2021,gomesQuasilocalDegreesFreedom2021}, i.e. the question of how the gauge-invariant data of two gauge subsystems can be combined to form the gauge-invariant data of the combined system. Gomes has answered this question as follows \cite[Theorem 1]{gomesHolismEmpiricalSignificance2021}.
\\
\begin{theorem}[Rigid variety for U(1)]
    For electromagnetism as coupled to a Klein-Gordon scalar field in a simply-connected universe: given the physical content of two
regions, for matter vanishing at the boundary but not in the bulk of the regions,
the universal state is underdetermined, resulting in a residual variety parametrized by
an element of U(1), where the particular action of U(1) is that which leaves the gauge-fields invariant, but
not the matter fields.
\end{theorem}
~\\
When this result is applied to the Abelian Higgs model it can be used to understand the physical origin of the Josephson effect. Considered individually, the global $U(1)$ phase of the Ginzburg-Landau order parameter of a superconductor is physically meaningless. But when a second superconductor is brought in, there will inevitably be interactions, implying that the two superconductors must be viewed as a combined system. The ambiguity in this combination is precisely the global $U(1)$ phase difference of the two Ginzburg-Landau order parameters. The Josephson current flows until this phase difference has been eliminated. In terms of Galileo's ship, this would be similar to bringing two separate ships moving at different speeds into some kind of contact, such that their relative velocity decreases until they move with the same speed. The relative velocity of the ships becomes a physically meaningful quantity from the moment any kind of interaction is established.

Thus, like in the case of Minkowski spacetime, there is a residual group of constant global gauge transformations acting also on the boundary of the superconductor, which carries direct, though relational, empirical significance. Different directions in which the $U(1)$ symmetry is broken can be thought of as superselection sectors. Changing the GL phase can only be achieved by external means, namely by adding another superconductor and using the Josephson current, or more generally by using any kind of coupling to the order parameter. In this sense the different ways to break the $U(1)$ symmetry correspond to sectors that cannot ``communicate'': one has to move out of the system to achieve a transition between them, analogously to how, for asymptotic boundaries, one has to perform an operation at infinity to change between superselection sectors.

Recognizing this, it becomes straightforward to relate the theory of asymptotic symmetry groups to bounded systems with finite boundaries. Indeed, one can think of the asymptotic boundary as representing a domain in which there is absolutely no external influence on the physical system in question, i.e. representing perfect isolation. The ``environment'' is then located at infinity. For finite systems this environment is finitely far away, and the boundary is an actual finite boundary. Then, one imposes the condition that the order parameter (in this case: the Higgs field) vanishes beyond the boundary to represent the idea that this boundary \textit{separates} the system from its (laboratory) environment, such that external perturbations do not affect the system. One then finds the same superselection structure for the finite system as for the infinite system, with the sectors labeled by all possible relational values that the corresponding order parameter could have with respect to the environment once the idealization of perfect separation is removed. But for the infinite system these sectors look intrinsic to the system, prompting us to call symmetry breaking \textit{spontaneous}. This is, however, an illusion that results from us having located the boundary at infinity. For a finite system it is immediately clear that the meaning of the different sectors lies in the relation with the environment. When the order parameter is required to vanish on the boundary, the system can remain in a symmetric state. But when it is recognized that this is an unrealistic idealization, and that really there will be sufficiently strong perturbations from the environment, the order parameter can no longer realistically be required to be strictly zero on the boundary and one finds a symmetry-broken ground state, induced ultimately by the interaction with the environment. In this respect, our view of SSB might align better with a so-called \textit{open systems view} \cite[]{Cuffaro}.

\section{Conclusion}\label{conclusion}

The enigmatic phenomenon of spontaneous symmetry breaking (SSB) continues to present profound conceptual challenges. Conventional definitions of SSB predominantly rely on the idealization of infinite quantum systems, a reliance that introduces significant tension when attempting to explain observed symmetry breaking in finite, real-world systems like a common fridge magnet or a superconductor. In this paper we have argued that boundary conditions are crucial for bridging the gap between SSB in infinite and finite systems. Though this is well-known for classical spin systems, especially in the context of Gibbs measures, we have applied this idea in a much wider sense and framed it explicitly in terms of the modern philosophical literature on symmetry. 

Our central proposal offers a novel perspective on SSB that shifts the focus to the interplay between a system and its environment at a boundary. We propose that a separating boundary condition, with the order parameter required to vanish on the boundary, creates superselection sectors because there is an ambiguity in how the system relates to the environment. This perspective is inherently relational, meaning that a system's symmetry is broken only in relation to an external reference across a boundary. For infinite systems this boundary is located ``at infinity'' and the superselection structure appears intrinsic to the system. For finite systems, however, perturbations from the environment will inevitably make the assumption of a vanishing order parameter on the boundary unrealistic, forcing the system into one superselection sector. The sector here indicates one of the possible stable configurations of the system with respect to its environment. This is quite similar to the description of classical phase transitions via Gibbs measures.

Our boundary-centered language should, however, be understood with some flexibility. For short-range interacting lattice systems, boundary conditions select a sector because the system has a spatial exterior: the environment can be idealized as acting at the edge of the system, or at infinity. In mean-field models such as the Curie-Weiss model, by contrast, sector selection is not mediated by a spatial boundary, since every degree of freedom couples to every other. The role of the environment is then better represented by a separate, macroscopic mean field, or by an arbitrarily small symmetry-breaking perturbation coupled directly to the order parameter as a whole. Still, the same conceptual structure survives: the symmetric finite state becomes unstable in the thermodynamic limit, extremal symmetry-broken states appear, and a symmetry-breaking reference selects one sector. Thus, in non-local models, the relevant ``boundary condition'' is not literally spatial, but is still needed in the specification of how the system's mean field relates to the mean field of its environment.

However, our account of SSB focuses on the problem of defining SSB kinematically and does not treat its \textit{dynamics}
\cite[]{landsmanAlgebraicTheorySuperselection1991, vanwezelSpontaneousSymmetryBreaking2008, vanwezel2010broken, landsmanFleaSchrodingerCat2013,borsboomspontaneity}.
The standard account of SSB is an equilibrium description; it describes the structure of the state space but provides no dynamical account of how a particular asymmetric ground state is selected. It is this standard account that we have attempted to improve in terms of superselection and boundary conditions. But even if one grants the existence of a symmetry-related superselection structure, the symmetric laws of motion provide no mechanism to propel the system from a symmetric state into a symmetry-broken sector, leaving a significant explanatory gap in the physical account \cite[]{landsmanSpontaneousSymmetryBreaking2013}. 

All in all, our proposal opens several intriguing questions and avenues for future research. 
How does this perspective, emphasizing the role of the environment and boundary conditions, intersect with the long-standing measurement problem in quantum mechanics? 
Could the act of measurement itself be understood as imposing specific boundary conditions that ``collapse'' a symmetric superposition into a broken-symmetry state? Furthermore, can this framework be rigorously extended to non-Abelian gauge theories or gravity? Finally, what are the direct experimental implications of viewing SSB in this relational way? Could designing experiments with precise control over boundary conditions allow for the observation or even manipulation of these ``boundary symmetries" and their breaking in novel ways? Exploring these questions will undoubtedly deepen our understanding of one of physics' most paradoxical phenomena.

\backmatter

\bmhead{Acknowledgments}

The authors want to thank Klaas Landsman and Jasper van Wezel for their guidance and elaborate feedback. The work of SB is supported by the Spinoza Grant of the Netherlands Organization for Scientific Research (NWO) awarded to Klaas Landsman and the work of JD by the Netherlands Organization for Scientific Research (NWO/OCW), as part of Quantum Limits (project number SUMMIT.1.1016).


\begin{appendices}




\end{appendices}


\bibliography{sn-bibliography}

@article{earmanCuriePrincipleSpontaneous2004,
  title = {Curie's {{Principle}} and Spontaneous Symmetry Breaking},
  author = {Earman, John},
  year = {2004},
  journal = {International Studies in the Philosophy of Science},
  volume = {18},
  number = {2-3},
  pages = {173--198},
  publisher = {{Routledge}},
  doi = {10.1080/0269859042000311299}
}

@article{fraserSSB,
         journal = {Philosophy of Science},
             doi = {10.1086/687263},
       publisher = {University of Chicago Press},
            year = {2016},
          volume = {83},
            note = {{\copyright}  2016 by the Philosophy of Science Association. All rights reserved. Reproduced in accordance with the publisher's self-archiving policy.},
           title = {Spontaneous Symmetry Breaking in Finite Systems},
           pages = {585--605},
          number = {4},
          author = {Fraser, JD},
            issn = {0031-8248},
             url = {http://doi.org/10.1086/687263}
}

@article{Wallace:2018zbu,
    author = "Wallace, David",
    title = "{Spontaneous Symmetry Breaking in Finite Quantum Systems: a decoherent-histories approach}",
    eprint = "1808.09547",
    archivePrefix = "arXiv",
    primaryClass = "quant-ph",
    month = "8",
    year = "2018"
}

@article{Donnelly:2016auv,
    author = "Donnelly, William and Freidel, Laurent",
    title = "{Local subsystems in gauge theory and gravity}",
    primaryClass = "hep-th",
    doi = "10.1007/JHEP09(2016)102",
    journal = "JHEP",
    volume = "09",
    pages = "102",
    year = "2016"
}

@article{Borsboom:2025agn,
    author = "Borsboom, Silvester and Posthuma, Hessel",
    title = "{Global Gauge Symmetries and Spatial Asymptotic Boundary Conditions in Yang-Mills theory}",
    eprint = "2502.16151",
    archivePrefix = "arXiv",
    primaryClass = "math-ph",
    year = "2025"
}

@book{binz,
  title = {Geometry of Classical Fields},
  author = {Binz, Ernst and Śniatycki, Jędrzej and Fischer, Hans},
  year = {1988},
  publisher = {Elsevier},
series = {North-Holland Mathematics Studies},
volume = {154}
}

@article{rielloHamiltonianGaugeTheory2024,
	title = {Hamiltonian gauge theory with corners: constraint reduction and flux superselection},
	volume = {28},
	doi = {10.4310/ATMP.241029014101},
	shorttitle = {Hamiltonian gauge theory with corners},
	pages = {1241--1424},
	number = {4},
	journal = {Advances in Theoretical and Mathematical Physics},
	author = {Riello, Aldo and Schiavina, Michele},
	year = {2024},
}

@article{sachsAsymptoticSymmetriesGravitational1962,
	title = {Asymptotic Symmetries in Gravitational Theory},
	volume = {128},
	url = {https://link.aps.org/doi/10.1103/PhysRev.128.2851},
	doi = {10.1103/PhysRev.128.2851},
	pages = {2851--2864},
	number = {6},
	journal = {Physical Review},
	shortjournal = {Phys. Rev.},
	author = {Sachs, R.},
	year = {1962},
	publisher = {American Physical Society}
}

@article{bondiGravitationalWavesGeneral1962,
	title = {Gravitational Waves in General Relativity. {VII}. Waves from Axi-Symmetric Isolated Systems},
	volume = {269},
	issn = {0080-46301364-5021},
	url = {https://ui.adsabs.harvard.edu/abs/1962RSPSA.269...21B},
	doi = {10.1098/rspa.1962.0161},
	journal = {Proceedings of the Royal Society of London Series A},
	author = {Bondi, H. and van der Burg, M. G. J. and Metzner, A. W. K.},
	year = {1962}
}

@article{SNIATYCKI1988291,
title = {Gauge invariance, boundary conditions, and charges},
journal = {Reports on Mathematical Physics},
volume = {25},
number = {3},
pages = {291-303},
year = {1988},
doi = {10.1016/0034-4877(88)90033-X},
author = {Jȩdrzej Sniatycki}
}

@inbook{landsmanrigour,
    author = {Klaas Landsman},
    title = {The Philosophy of Rigour},
    publisher = {Routledge},
    year = {2025},
    note = {Forthcoming},
    chapter = {Rigour from rules: Deduction and definition in mathematical physics},
    editor = {Dean Rickles and Karim Thébault}
}

@incollection{tasaki2020long,
  title={Long-Range Order and Spontaneous Symmetry Breaking in the Classical and Quantum Ising Models},
  author={Tasaki, Hal},
  booktitle={Physics and Mathematics of Quantum Many-Body Systems},
  pages={49--71},
  year={2020},
  publisher={Springer}
}

@article{vanwezelSpontaneousSymmetryBreaking2008,
  title = {Spontaneous Symmetry Breaking and Decoherence in Superconductors},
  author = {Wezel, Jasper van and Brink, Jeroen van den},
  year = {2008},
  journal = {Physical Review B},
  shortjournal = {Phys. Rev. B},
  volume = {77},
  number = {6},
  pages = {064523},
  publisher = {American Physical Society},
  doi = {10.1103/PhysRevB.77.064523}
}

@article{Gotay:1997eg,
    author = "Gotay, Mark J. and Isenberg, James and Marsden, Jerrold E.",
    title = "{Momentum maps and classical relativistic fields. Part 1: Covariant Field Theory}",
    eprint = "physics/9801019",
    archivePrefix = "arXiv",
    month = "11",
    year = "1997"
}

@article{Gotay:2004ib,
    author = "Gotay, Mark J. and Isenberg, James and Marsden, Jerrold E.",
    title = "{Momentum maps and classical relativistic fields. Part II: Canonical analysis of field theories}",
    eprint = "math-ph/0411032",
    archivePrefix = "arXiv",
    month = "8",
    year = "2004"
}

@inproceedings{Gotay2006MomentumMA,
  title={Momentum Maps and Classical Fields Part III: Gauge Symmetries and Initial Value Constraints},
  author={Mark J. Gotay and Jerrold E. Marsden},
  year={2006},
  url={https://api.semanticscholar.org/CorpusID:214797075}
}

@InProceedings{sniatycskiboundarycondspatboundeddom,
author="{Sniatycki}, J{\k{e}}drzej",
editor="Hennig, J{\"o}-Dieter
and L{\"u}cke, Wolfgang
and Tolar, Ji{\v{r}}{\'i}",
title="On boundary conditions for {Yang}-{Mills} fields in spatially bounded domains",
booktitle="Differential Geometry, Group Representations, and Quantization",
year="1991",
publisher="Springer Berlin Heidelberg",
address="Berlin, Heidelberg",
pages="43--53",
doi ={10.1007/3-540-53941-7_3}
}

@misc{borsboomspontaneity,
          author = {Silvester Borsboom},
            year = {2024},
           title = {Spontaneity in {Nature} and its {Relation} to {Randomness} and {Indeterminism}},
            note = {Available at \url{https://philsci-archive.pitt.edu/26007/}}
}

@article{stromingerAsymptoticSymmetriesYangMills2014,
	title = {Asymptotic symmetries of {Yang}-{Mills} theory},
	volume = {2014},
	issn = {1029-8479},
	url = {https://doi.org/10.1007/JHEP07(2014)151},
	doi = {10.1007/JHEP07(2014)151},
	pages = {151},
	number = {7},
	journal = {Journal of High Energy Physics},
	shortjournal = {J. High Energ. Phys.},
	author = {Strominger, Andrew},
	urldate = {2025-01-23},
	year = {2014},
	langid = {english},
}

@inproceedings{pasterskiCelestialHolography2021,
    author = "Pasterski, Sabrina and Pate, Monica and Raclariu, Ana-Maria",
    title = "{Celestial Holography}",
    booktitle = "{Snowmass 2021}",
    eprint = "2111.11392",
    archivePrefix = "arXiv",
    primaryClass = "hep-th",
    month = "11",
    year = "2021"
}

@article{rielloEdgeModesEdge2021,
    author = "Riello, Aldo",
    title = "{Edge modes without edge modes}",
    eprint = "2104.10182",
    archivePrefix = "arXiv",
    primaryClass = "hep-th",
    month = "4",
    year = "2021"
}

@misc{stromingerLecturesInfraredStructure2018,
	title = {Lectures on the Infrared Structure of Gravity and Gauge Theory},
	url = {http://arxiv.org/abs/1703.05448},
	doi = {10.48550/arXiv.1703.05448},
	number = {{arXiv}:1703.05448},
	publisher = {{arXiv}},
	author = {Strominger, Andrew},
	urldate = {2025-02-17},
	year = {2018},
	eprinttype = {arxiv},
	eprint = {1703.05448 [hep-th]},
note = {Available at arXiv:1703.05448, \url{http://arxiv.org/abs/1703.05448}},
}

@article{henneauxAsymptoticSymmetriesElectromagnetism2018,
	title = {Asymptotic symmetries of electromagnetism at spatial infinity},
	volume = {2018},
	issn = {1029-8479},
	url = {https://doi.org/10.1007/JHEP05(2018)137},
	doi = {10.1007/JHEP05(2018)137},
	pages = {137},
	number = {5},
	journal = {Journal of High Energy Physics},
	shortjournal = {J. High Energ. Phys.},
	author = {Henneaux, Marc and Troessaert, Cédric},
	urldate = {2025-01-24},
	year = {2018},
	langid = {english},
}

@article{Tanzi:2020fmt,
    author = "Tanzi, Roberto and Giulini, Domenico",
    title = "{Asymptotic symmetries of Yang-Mills fields in Hamiltonian formulation}",
    doi = "10.1007/JHEP10(2020)094",
    journal = "JHEP",
    volume = "10",
    pages = "094",
    year = "2020"
}

@article{Giulini:1994bi,
    author = "Giulini, Domenico",
    title = "{Asymptotic symmetry groups of long ranged gauge configurations}",
    eprint = "gr-qc/9410042",
    archivePrefix = "arXiv",
    reportNumber = "CGPG-94-10-5",
    doi = "10.1142/S0217732395002210",
    journal = "Mod. Phys. Lett. A",
    volume = "10",
    pages = "2059--2070",
    year = "1995"
}

@article{rielloNullHamiltonianYang2025,
	title = {Null Hamiltonian Yang–Mills theory: Soft Symmetries and Memory as Superselection},
	volume = {26},
	doi = {10.1007/s00023-024-01428-z},
	shorttitle = {Null Hamiltonian Yang–Mills theory},
	pages = {389--477},
	number = {2},
	journal = {Annales Henri Poincaré},
	shortjournal = {Ann. Henri Poincaré},
	author = {Riello, Aldo and Schiavina, Michele},
	year = {2025},
	langid = {english},
}

@book{landsmanFoundationsQuantumTheory2017,
  title = {Foundations of {{Quantum Theory}}},
  author = {Landsman, Klaas},
  year = {2017},
  series = {Fundamental {{Theories}} of {{Physics}}},
  volume = {188},
  publisher = {Springer International Publishing},
  location = {Cham}
}

@article{fraserHiggsMechanismSuperconductivity2016,
  title = {The {{Higgs}} Mechanism and Superconductivity: {{A}} Case Study of Formal Analogies},
  shorttitle = {The {{Higgs}} Mechanism and Superconductivity},
  author = {Fraser, Doreen and Koberinski, Adam},
  year = {2016},
  journal = {Studies in History and Philosophy of Science Part B: Studies in History and Philosophy of Modern Physics},
  shortjournal = {Studies in History and Philosophy of Science Part B: Studies in History and Philosophy of Modern Physics},
  volume = {55},
  pages = {72--91},
  issn = {1355-2198},
  doi = {10.1016/j.shpsb.2016.08.003},
  url = {https://www.sciencedirect.com/science/article/pii/S1355219816300739},
  urldate = {2023-03-22},
  langid = {english}
}

@article{nambuDynamicalModelElementary1961a,
  title = {Dynamical {{Model}} of {{Elementary Particles Based}} on an {{Analogy}} with {{Superconductivity}}. {{II}}},
  author = {Nambu, Y. and Jona-Lasinio, G.},
  year = {1961},
  journal = {Physical Review},
  shortjournal = {Phys. Rev.},
  volume = {124},
  number = {1},
  pages = {246--254},
  publisher = {{American Physical Society}},
  doi = {10.1103/PhysRev.124.246},
  url = {https://link.aps.org/doi/10.1103/PhysRev.124.246},
  urldate = {2023-06-01}
}

@article{nambuDynamicalModelElementary1961,
  title = {Dynamical {{Model}} of {{Elementary Particles Based}} on an {{Analogy}} with {{Superconductivity}}. {{I}}},
  author = {Nambu, Y. and Jona-Lasinio, G.},
  year = {1961},
  journal = {Physical Review},
  shortjournal = {Phys. Rev.},
  volume = {122},
  number = {1},
  pages = {345--358},
  publisher = {{American Physical Society}},
  doi = {10.1103/PhysRev.122.345},
  url = {https://link.aps.org/doi/10.1103/PhysRev.122.345},
  urldate = {2023-06-01}
}

@article{higgsBrokenSymmetriesMasses1964,
  title = {Broken {{Symmetries}} and the {{Masses}} of {{Gauge Bosons}}},
  author = {Higgs, Peter W.},
  year = {1964},
  journal = {Physical Review Letters},
  shortjournal = {Phys. Rev. Lett.},
  volume = {13},
  number = {16},
  pages = {508--509},
  publisher = {{American Physical Society}},
  doi = {10.1103/PhysRevLett.13.508},
  url = {https://link.aps.org/doi/10.1103/PhysRevLett.13.508},
  urldate = {2023-06-02}
}

@article{Higgs:1964ia,
    author = "Higgs, Peter W.",
    title = "{Broken symmetries, massless particles and gauge fields}",
    doi = "10.1016/0031-9163(64)91136-9",
    journal = "Phys. Lett.",
    volume = "12",
    pages = "132--133",
    year = "1964"
}

@article{englertBrokenSymmetryMass1964,
  title = {Broken {{Symmetry}} and the {{Mass}} of {{Gauge Vector Mesons}}},
  author = {Englert, F. and Brout, R.},
  year = {1964},
  journal = {Physical Review Letters},
  shortjournal = {Phys. Rev. Lett.},
  volume = {13},
  number = {9},
  pages = {321--323},
  publisher = {{American Physical Society}},
  doi = {10.1103/PhysRevLett.13.321},
  url = {https://link.aps.org/doi/10.1103/PhysRevLett.13.321},
  urldate = {2023-06-02}
}

@article{guralnikGlobalConservationLaws1964,
  title = {Global {{Conservation Laws}} and {{Massless Particles}}},
  author = {Guralnik, G. S. and Hagen, C. R. and Kibble, T. W. B.},
  year = {1964},
  journal = {Physical Review Letters},
  shortjournal = {Phys. Rev. Lett.},
  volume = {13},
  number = {20},
  pages = {585--587},
  publisher = {{American Physical Society}},
  doi = {10.1103/PhysRevLett.13.585},
  url = {https://link.aps.org/doi/10.1103/PhysRevLett.13.585},
  urldate = {2023-06-02}
}

@article{earmanLawsSymmetrySymmetry2004,
  title = {Laws, {{Symmetry}}, and {{Symmetry Breaking}}: {{Invariance}}, {{Conservation Principles}}, and {{Objectivity}}},
  shorttitle = {Laws, {{Symmetry}}, and {{Symmetry Breaking}}},
  author = {Earman, John},
  year = {2004},
  journal = {Philosophy of Science},
  volume = {71},
  number = {5},
  pages = {1227--1241},
  publisher = {{Cambridge University Press}},
  issn = {0031-8248, 1539-767X},
  doi = {10.1086/428016},
  url = {https://www.cambridge.org/core/journals/philosophy-of-science/article/abs/laws-symmetry-and-symmetry-breaking-invariance-conservation-principles-and-objectivity/0BC63A7655798CFA8ED1D533F9D0DAB6#},
  urldate = {2023-08-17},
  langid = {english}
}

@article{lyreDoesHiggsMechanism2008,
  title = {Does the {{Higgs Mechanism Exist}}?},
  author = {Lyre, Holger},
  year = {2008},
  journal = {International Studies in the Philosophy of Science},
  volume = {22},
  number = {2},
  pages = {119--133},
  publisher = {{Routledge}},
  issn = {0269-8595},
  doi = {10.1080/02698590802496664},
  url = {https://doi.org/10.1080/02698590802496664},
  urldate = {2023-08-28}
}

@article{Ginzburg:1950sr,
    author = "Ginzburg, V. L. and Landau, L. D.",
    editor = "ter Haar, D.",
    title = "{On the Theory of superconductivity}",
    doi = "10.1016/b978-0-08-010586-4.50078-x",
    journal = "Zh. Eksp. Teor. Fiz.",
    volume = "20",
    pages = "1064--1082",
    year = "1950"
}

@article{smeenkElusiveHiggsMechanism2006,
  title = {The {{Elusive Higgs Mechanism}}},
  author = {Smeenk, Chris},
  year = {2006},
  journal = {Philosophy of Science},
  volume = {73},
  number = {5},
  pages = {487--499},
  publisher = {{[The University of Chicago Press, Philosophy of Science Association]}},
  issn = {0031-8248},
  doi = {10.1086/518324}
}

@article{struyveGaugeInvariantAccounts2011a,
  title = {Gauge Invariant Accounts of the {{Higgs}} Mechanism},
  author = {Struyve, Ward},
  year = {2011},
  journal = {Studies in History and Philosophy of Modern Physics},
  volume = {42},
  number = {4},
  pages = {226--236},
  publisher = {{Elsevier}},
  doi ={10.1016/j.shpsb.2011.06.003}
}

@article{friederichGaugeSymmetryBreaking2013,
  title = {Gauge Symmetry Breaking in Gauge Theories—in Search of Clarification},
  author = {Friederich, Simon},
  year = {2013},
  journal = {European Journal for Philosophy of Science},
  shortjournal = {Euro Jnl Phil Sci},
  volume = {3},
  number = {2},
  pages = {157--182},
  issn = {1879-4920},
  doi = {10.1007/s13194-012-0061-y},
  url = {https://doi.org/10.1007/s13194-012-0061-y},
  urldate = {2023-09-14},
  langid = {english}
}

@book{haagLocalQuantumPhysics1996,
  title = {Local {{Quantum Physics}}},
  author = {Haag, Rudolf},
  year = {1996},
  publisher = {{Springer}},
  location = {{Berlin, Heidelberg}},
  doi = {10.1007/978-3-642-61458-3},
  isbn = {978-3-540-61049-6 978-3-642-61458-3}
}

@book{berghoferGaugeSymmetriesSymmetry2023,
  title = {Gauge {{Symmetries}}, {{Symmetry Breaking}}, and {{Gauge-Invariant Approaches}}},
  author = {Berghofer, Philipp and François, Jordan and Friederich, Simon and Gomes, Henrique and Hetzroni, Guy and Maas, Axel and Sondenheimer, René},
  year = {2023},
  publisher = {{Cambridge University Press}},
  doi = {10.1017/9781009197236}
}

@article{josephsonPossibleNewEffects1962,
  title = {Possible New Effects in Superconductive Tunnelling},
  author = {Josephson, B. D.},
  year = {1962},
  journal = {Physics Letters},
  shortjournal = {Physics Letters},
  volume = {1},
  number = {7},
  pages = {251--253},
  issn = {0031-9163},
  doi = {10.1016/0031-9163(62)91369-0},
  url = {https://www.sciencedirect.com/science/article/pii/0031916362913690},
  urldate = {2023-12-06}
}

@article{morchioLocalizationSymmetries2007,
  title = {Localization and Symmetries},
  author = {Morchio, G. and Strocchi, F.},
  year = {2007},
  journal = {Journal of Physics A: Mathematical and Theoretical},
  shortjournal = {J. Phys. A: Math. Theor.},
  volume = {40},
  number = {12},
  pages = {3173},
  issn = {1751-8121},
  doi = {10.1088/1751-8113/40/12/S17},
  url = {https://dx.doi.org/10.1088/1751-8113/40/12/S17},
  urldate = {2023-12-07},
  langid = {english}
}

@article{vanwezel2007spontaneous,
  title   = {Spontaneous symmetry breaking in quantum mechanics},
  author  = {{Wezel}, Jasper van and Brink, Jeroen van den},
  journal = {American Journal of Physics},
  volume  = {75},
  number  = {7},
  pages   = {635--638},
  year    = {2007},
  doi     = {10.1119/1.2730839}
}

@article{vanwezel2010broken,
  title={Broken time translation symmetry as a model for quantum state reduction},
  author={Wezel, Jasper van},
  journal={Symmetry},
  volume={2},
  number={2},
  pages={582--608},
  year={2010},
  publisher={Molecular Diversity Preservation International},
  doi ={10.3390/sym2020582}
}

@book{strocchiSymmetryBreaking2008,
  title = {Symmetry {{Breaking}}},
  author = {Strocchi, Franco},
  year = {2008},
  series = {Lecture {{Notes}} in {{Physics}}},
  volume = {732},
  publisher = {{Springer}},
  location = {{Berlin, Heidelberg}},
  doi = {10.1007/978-3-540-73593-9},
  url = {http://link.springer.com/10.1007/978-3-540-73593-9},
  urldate = {2023-12-07},
  isbn = {978-3-540-73592-2 978-3-540-73593-9},
  langid = {english}
}

@book{strocchiSymmetryBreakingStandard2019,
  title = {Symmetry {{Breaking}} in the {{Standard Model}}: {{A Non-Perturbative Outlook}}},
  shorttitle = {Symmetry {{Breaking}} in the {{Standard Model}}},
  author = {Strocchi, Franco},
  year = {2019},
  publisher = {{Scuola Normale Superiore}},
  location = {{Pisa}},
  doi = {10.1007/978-88-7642-660-5},
  url = {http://link.springer.com/10.1007/978-88-7642-660-5},
  urldate = {2023-12-07},
  isbn = {978-88-7642-659-9 978-88-7642-660-5},
  langid = {english}
}

@book{strocchiIntroductionNonPerturbativeFoundations2016,
  title = {An {{Introduction}} to {{Non-Perturbative Foundations}} of {{Quantum Field Theory}}},
  author = {Strocchi, Franco},
  year = {2013},
  series = {International {{Series}} of {{Monographs}} on {{Physics}}},
  publisher = {{Oxford University Press}},
  location = {{Oxford, New York}},
  isbn = {978-0-19-878923-9},
  pagetotal = {270}
}

@article{greavesEmpiricalConsequencesSymmetries2014,
  author  = {Greaves, Hilary and Wallace, David},
  title   = {Empirical Consequences of Symmetries},
  journal = {The British Journal for the Philosophy of Science},
  volume  = {65},
  number  = {1},
  pages   = {59--89},
  year    = {2014},
  doi     = {10.1093/bjps/axt005}
}

@article{lusannaDiracObservablesHiggs1997,
  title = {Dirac's {{Observables}} for the {{Higgs Model}}: {{I}}) the {{Abelian Case}}},
  shorttitle = {Dirac's {{Observables}} for the {{Higgs Model}}},
  author = {Lusanna, Luca and Valtancoli, Paolo},
  year = {1997},
  journal = {International Journal of Modern Physics A},
  shortjournal = {Int. J. Mod. Phys. A},
  volume = {12},
  number = {26},
doi = {10.1142/S0217751X97002541}
}

@book{henneauxQuantizationGaugeSystems1992,
  title = {Quantization of {{Gauge Systems}}},
  author = {Henneaux, Marc and Teitelboim, Claudio},
  year = {1992},
  eprint = {j.ctv10crg0r},
  eprinttype = {jstor},
  publisher = {{Princeton University Press}},
  doi = {10.2307/j.ctv10crg0r},
  url = {https://www.jstor.org/stable/j.ctv10crg0r},
  urldate = {2024-01-17},
  isbn = {978-0-691-08775-7}
}

@article{gomesGaugingBoundaryFieldspace2019,
  title = {Gauging the Boundary in Field-Space},
  author = {Gomes, Henrique},
  year = {2019},
  journal = {Studies in History and Philosophy of Science Part B: Studies in History and Philosophy of Modern Physics},
  shortjournal = {Studies in History and Philosophy of Science Part B: Studies in History and Philosophy of Modern Physics},
  volume = {67},
  pages = {89--110},
  issn = {1355-2198},
  doi = {10.1016/j.shpsb.2019.04.002},
  url = {https://www.sciencedirect.com/science/article/pii/S1355219818302144},
  urldate = {2024-02-08}
}

@article{gomesHolismEmpiricalSignificance2021,
  title = {Holism as the Empirical Significance of Symmetries},
  author = {Gomes, Henrique},
  year = {2021},
  journal = {European Journal for Philosophy of Science},
  shortjournal = {Euro Jnl Phil Sci},
  volume = {11},
  number = {3},
  pages = {87},
  issn = {1879-4920},
  doi = {10.1007/s13194-021-00397-y},
  url = {https://doi.org/10.1007/s13194-021-00397-y},
  urldate = {2024-02-08},
  langid = {english}
}

@article{gomesQuasilocalDegreesFreedom2021,
  title = {The Quasilocal Degrees of Freedom of {{Yang-Mills}} Theory},
  author = {Gomes, Henrique and Riello, Aldo},
  year = {2021},
  journal = {SciPost Physics},
  volume = {10},
  number = {6},
  pages = {130},
  issn = {2542-4653},
  doi = {10.21468/SciPostPhys.10.6.130},
  url = {https://scipost.org/SciPostPhys.10.6.130},
  urldate = {2024-02-08},
  langid = {english}
}

@article{gomesUnifiedGeometricFramework2018,
  title = {Unified Geometric Framework for Boundary Charges and Particle Dressings},
  author = {Gomes, Henrique and Riello, Aldo},
  year = {2018},
  journal = {Physical Review D},
  shortjournal = {Phys. Rev. D},
  volume = {98},
  number = {2},
  pages = {025013},
  publisher = {{American Physical Society}},
  doi = {10.1103/PhysRevD.98.025013},
  url = {https://link.aps.org/doi/10.1103/PhysRevD.98.025013},
  urldate = {2024-02-08}
}

@article{tehGalileoGaugeUnderstanding2016,
  title = {Galileo’s {{Gauge}}: {{Understanding}} the {{Empirical Significance}} of {{Gauge Symmetry}}},
  shorttitle = {Galileo’s {{Gauge}}},
  author = {Teh, Nicholas J.},
  year = {2016},
  journal = {Philosophy of Science},
  volume = {83},
  number = {1},
  pages = {93--118},
  publisher = {{Cambridge University Press}},
  issn = {0031-8248, 1539-767X},
  doi = {10.1086/684196},
  url = {https://www.cambridge.org/core/journals/philosophy-of-science/article/galileos-gauge-understanding-the-empirical-significance-of-gauge-symmetry/9C32EDE0DDBC354E13A46C1289AC8571},
  urldate = {2024-02-14},
  langid = {english}
}

@article{wallaceIsolatedSystemsTheireen,
  title = {Isolated Systems and Their Symmetries, Part {{I}}: {{General}} Framework and Particle-Mechanics Examples},
  author = {Wallace, David},
  year = {2022},
  journal = {Studies in History and Philosophy of Science},
  volume = {92},
  pages = {239--248},
  issn = {0039-3681},
  doi = {10.1016/j.shpsa.2022.01.015},
  url = {https://www.sciencedirect.com/science/article/pii/S0039368122000255}
}

@article{wallaceIsolatedSystemstwee,
  title = {Isolated {{Systems}} and {{Their Symmetries}}, {{Part II}}: {{Local}} and {{Global Symmetries}} of {{Field Theories}}},
  shorttitle = {Isolated {{Systems}} and {{Their Symmetries}}, {{Part Ii}}},
  author = {Wallace, David},
  year = {2022},
  journal = {Studies in History and Philosophy of Science Part A},
  volume = {92},
  pages = {249--259},
  doi = {10.1016/j.shpsa.2022.01.016}
}

@article{gomesUnifiedGeometricFramework2019,
  title = {A Unified Geometric Framework for Boundary Charges and Dressings: {{Non-Abelian}} Theory and Matter},
  shorttitle = {A Unified Geometric Framework for Boundary Charges and Dressings},
  author = {Gomes, Henrique and Hopfmüller, Florian and Riello, Aldo},
  year = {2019},
  journal = {Nuclear Physics B},
  shortjournal = {Nuclear Physics B},
  volume = {941},
  pages = {249--315},
  issn = {0550-3213},
  doi = {10.1016/j.nuclphysb.2019.02.020},
  url = {https://www.sciencedirect.com/science/article/pii/S0550321319300483},
  urldate = {2024-02-16}
}

@book{ruetscheInterpretingQuantumTheories2011,
  title = {Interpreting {{Quantum Theories}}: {{The Art}} of the {{Possible}}},
  shorttitle = {Interpreting {{Quantum Theories}}},
  author = {Ruetsche, Laura},
  year = {2011},
  publisher = {Oxford University Press UK},
  location = {Oxford, GB}
}

@article{landsmanAlgebraicTheorySuperselection1991,
  title = {Algebraic Theory of Superselection Sectors and the Measurement Problem in Quantum Mechanics},
  author = {Landsman, N. P.},
  year = {1991},
  journal = {International Journal of Modern Physics A},
  shortjournal = {Int. J. Mod. Phys. A},
  volume = {06},
  number = {30},
  pages = {5349--5371},
  publisher = {World Scientific Publishing Co.},
  doi ={10.1142/S0217751X91002513}
}

@misc{landsmanClassicalQuantum2005,
  title = {Between Classical and Quantum},
  author = {Landsman, N. P.},
  year = {2005},
  note = {Preprint available at arXiv:quant-ph/0506082, \url{http://arxiv.org/abs/quant-ph/0506082}},
  eprint = {quant-ph/0506082},
  eprinttype = {arxiv},
  doi = {10.48550/arXiv.quant-ph/0506082},
  url = {http://arxiv.org/abs/quant-ph/0506082},
  urldate = {2024-03-12},
  pubstate = {preprint}
}

@article{earmanSuperselectionRulesPhilosophers2008,
  title = {Superselection {{Rules}} for {{Philosophers}}},
  author = {Earman, John},
  year = {2008},
  journal = {Erkenntnis (1975-)},
  volume = {69},
  number = {3},
  pages = {377--414},
  publisher = {Springer},
  issn = {0165-0106},
  doi ={10.1007/s10670-008-9124-z}
}

@article{doplicherFieldsObservablesGauge1969,
  title = {Fields, Observables and Gauge Transformations {{I}}},
  author = {Doplicher, Sergio and Haag, Rudolf and Roberts, John E.},
  year = {1969},
  journal = {Communications in Mathematical Physics},
  shortjournal = {Commun.Math. Phys.},
  volume = {13},
  number = {1},
  pages = {1--23},
  issn = {1432-0916},
  doi = {10.1007/BF01645267},
  url = {https://doi.org/10.1007/BF01645267},
  urldate = {2024-03-13},
  langid = {english}
}

@article{doplicherFieldsObservablesGauge1969a,
  title = {Fields, Observables and Gauge Transformations {{II}}},
  author = {Doplicher, Sergio and Haag, Rudolf and Roberts, John E.},
  year = {1969},
  journal = {Communications in Mathematical Physics},
  shortjournal = {Commun.Math. Phys.},
  volume = {15},
  number = {3},
  pages = {173--200},
  issn = {1432-0916},
  doi = {10.1007/BF01645674},
  url = {https://doi.org/10.1007/BF01645674},
  urldate = {2024-03-13},
  langid = {english}
}

@article{doplicherWhyThereField1990,
  title = {Why There Is a Field Algebra with a Compact Gauge Group Describing the Superselection Structure in Particle Physics},
  author = {Doplicher, Sergio and Roberts, John E.},
  year = {1990},
  journal = {Communications in Mathematical Physics},
  shortjournal = {Commun.Math. Phys.},
  volume = {131},
  number = {1},
  pages = {51--107},
  issn = {1432-0916},
  doi = {10.1007/BF02097680},
  url = {https://doi.org/10.1007/BF02097680},
  urldate = {2024-03-13},
  langid = {english}
}

@article{doplicherLocalObservablesParticle1971,
  title = {Local Observables and Particle Statistics {{I}}},
  author = {Doplicher, Sergio and Haag, Rudolf and Roberts, John E.},
  year = {1971},
  journal = {Communications in Mathematical Physics},
  shortjournal = {Commun.Math. Phys.},
  volume = {23},
  number = {3},
  pages = {199--230},
  issn = {1432-0916},
  doi = {10.1007/BF01877742},
  url = {https://doi.org/10.1007/BF01877742},
  urldate = {2024-03-14},
  langid = {english}
}

@book{marsdenIntroductionMechanicsSymmetry1999,
  title = {Introduction to {{Mechanics}} and {{Symmetry}}: {{A Basic Exposition}} of {{Classical Mechanical Systems}}},
  shorttitle = {Introduction to {{Mechanics}} and {{Symmetry}}},
  author = {Marsden, Jerrold E. and Ratiu, Tudor S.},
  editorb = {Marsden, Jerrold E. and Sirovich, L. and Golubitsky, M. and Jäger, W.},
  editorbtype = {redactor},
  year = {1999},
  series = {Texts in {{Applied Mathematics}}},
  volume = {17},
  publisher = {Springer},
  location = {New York, NY},
  doi = {10.1007/978-0-387-21792-5},
  url = {http://link.springer.com/10.1007/978-0-387-21792-5},
  urldate = {2024-03-20},
  isbn = {978-1-4419-3143-6 978-0-387-21792-5},
  langid = {english}
}

@article{doplicherNewDualityTheory1989,
  title = {A New Duality Theory for Compact Groups},
  author = {Doplicher, Sergio and Roberts, John E.},
  year = {1989},
  journal = {Inventiones mathematicae},
  shortjournal = {Invent Math},
  volume = {98},
  number = {1},
  pages = {157--218},
  issn = {1432-1297},
  doi = {10.1007/BF01388849},
  url = {https://doi.org/10.1007/BF01388849},
  urldate = {2024-04-11},
  langid = {english}
}

@article{koma1994symmetry,
  title={Symmetry breaking and finite-size effects in quantum many-body systems},
  author={Koma, Tohru and Tasaki, Hal},
  journal={Journal of statistical physics},
  volume={76},
  number={3},
  pages={745--803},
  year={1994},
  publisher={Springer},
  doi ={10.1007/BF02188685}
}

@article{landsmanFleaSchrodingerCat2013,
  title = {A {{Flea}} on {{Schrödinger}}'s {{Cat}}},
  shorttitle = {A {{Flea}} on {{Schrödinger}}?},
  author = {Landsman, N. P. and Reuvers, Robin},
  year = {2013},
  journal = {Foundations of Physics},
  volume = {43},
  number = {3},
  pages = {373--407},
  publisher = {Springer US},
  doi = {10.1007/s10701-013-9700-1}
}

@article{Greiter2005,
  author        = {Greiter, Martin},
  title         = {Is electromagnetic gauge invariance spontaneously violated in superconductors?},
  journal       = {Annals of Physics},
  volume        = {319},
  number        = {1},
  pages         = {217--249},
  year          = {2005},
  doi           = {10.1016/j.aop.2005.03.008}
}

@article{kubo1957statistica,
    author = "Kubo, Ryogo",
    title = "{Statistical mechanical theory of irreversible processes. 1. General theory and simple applications in magnetic and conduction problems}",
    doi = "10.1143/JPSJ.12.570",
    journal = "J. Phys. Soc. Jap.",
    volume = "12",
    pages = "570--586",
    year = "1957"
}

@article{martin1959j,
  title={Theory of many-particle systems. I},
  author={Martin, Paul C and Schwinger, Julian},
  journal={Physical Review},
  volume={115},
  number={6},
  pages={1342-1373},
  year={1959},
  publisher={APS},
  doi ={10.1103/PhysRev.115.1342}
}

@book{suzuki2012quantum,
  title={Quantum Ising phases and transitions in transverse Ising models},
  author={Suzuki, Sei and Inoue, Jun-ichi and Chakrabarti, Bikas K},
  volume={862},
  year={2013},
  publisher={Springer}
}

@article{haag1967equilibrium,
  title={On the equilibrium states in quantum statistical mechanics},
  author={Haag, Rudolf and Hugenholtz, Nicolaas Marinus and Winnink, Marinus},
  journal={Communications in Mathematical Physics},
  volume={5},
  number={3},
  pages={215--236},
  year={1967},
  publisher={Springer},
  doi = {10.1007/BF01646342}
}

@article{anderson1952approximate,
    author = "Anderson, P. W.",
    title = "{An Approximate Quantum Theory of the Antiferromagnetic Ground State}",
    doi = "10.1103/PhysRev.86.694",
    journal = "Phys. Rev.",
    volume = "86",
    number = "5",
    pages = "694",
    year = "1952"
}

@article{coleman1987mixed,
  title = {Mixed valence as an almost broken symmetry},
  author = {Coleman, Piers},
  journal = {Phys. Rev. B},
  volume = {35},
  issue = {10},
  pages = {5072--5116},
  year = {1987},
  publisher = {American Physical Society},
  doi = {10.1103/PhysRevB.35.5072}
}

@article{bogoliubov1961quasi,
  title={Quasi-averages in problems of statistical mechanics},
  author={Bogoliubov, NN},
  journal={Preprint JINR D-781,(JINR, Dubna 1961)},
  year={1961}
}

@book{bratteli1998operator,
  title={Operator Algebras and Quantum Statistical Mechanics: Equilibrium States. Models in Quantum Statistical Mechanics},
  author={Bratteli, O. and Robinson, D.W.},
  isbn={9783540614432},
  lccn={gb97047698},
  series={Theoretical and Mathematical Physics},
  year={1998},
  publisher={Springer Berlin Heidelberg}
}

@article{pfeuty1970one,
  title={The one-dimensional Ising model with a transverse field},
  author={Pfeuty, Pierre},
  journal={ANNALS of Physics},
  volume={57},
  number={1},
  pages={79--90},
  year={1970},
  publisher={Elsevier},
  doi ={10.1016/0003-4916(70)90270-8}
}

@article{kogut1979introduction,
    author = "Kogut, John B.",
    title = "{An Introduction to Lattice Gauge Theory and Spin Systems}",
    reportNumber = "ILL-TH-79-4",
    doi = "10.1103/RevModPhys.51.659",
    journal = "Rev. Mod. Phys.",
    volume = "51",
    pages = "659",
    year = "1979"
}

@article{landsmanSpontaneousSymmetryBreaking2013,
  title = {Spontaneous {{Symmetry Breaking}} in {{Quantum Systems}}: {{Emergence}} or {{Reduction}}?},
  shorttitle = {Spontaneous {{Symmetry Breaking}} in {{Quantum Systems}}},
  author = {Landsman, Nicolaas P.},
  year = {2013},
  journal = {Studies in History and Philosophy of Science Part B: Studies in History and Philosophy of Modern Physics},
  volume = {44},
  number = {4},
  pages = {379--394},
  publisher = {Elsevier},
  doi = {10.1016/j.shpsb.2013.07.003}
}

@article{tehPhilosophySymmetry2024,
  title = {The {{Philosophy}} of {{Symmetry}}},
  author = {Teh, Nicholas Joshua Yii Wye},
  year = {2024},
  journal = {Elements in the Philosophy of Physics},
  series = {Cambridge {{Elements}}},
  publisher = {Cambridge University Press},
  doi = {10.1017/9781009008600},
  url = {https://www.cambridge.org/core/elements/philosophy-of-symmetry/0F4F206DF222DEE65E2314F578AFAEC6},
  urldate = {2024-06-13},
  isbn = {9781009008600 9781009507301 9781009005043},
  langid = {english}
}

@misc{borsboom2024spontaneous,
  title = {Spontaneous {Breaking} of {Global} {Gauge} {Symmetries} in the {Higgs} {Mechanism}},
  author = {Borsboom, Silvester},
  year = {2024},
  volume = {MSc thesis, University of Amsterdam},
  url = {https://scripties.uba.uva.nl/search?id=record_54629},
  urldate = {2024-09-04},
note = {Available at PhilSci Archive, \url{https://philsci-archive.pitt.edu/24403/}}
}

@article{borsboomdeharo,
  title = {Global {Gauge} {Symmetry} {Breaking} in the {Abelian} {Higgs} {Mechanism}},
  author = {Borsboom, Silvester and de Haro, Sebastian},
    eprint = "2504.17483",
    archivePrefix = "arXiv",
    primaryClass = "hist-ph",
    year = "2025"
}

@article{Rovelli:2013fga,
    author = "Rovelli, Carlo",
    title = "{Why Gauge?}",
    doi = "10.1007/s10701-013-9768-7",
    journal = "Found. Phys.",
    volume = "44",
    number = "1",
    pages = "91--104",
    year = "2014"
}

@article{Kabel:2024lzr,
    author = "Kabel, Viktoria and de la Hamette, Anne-Catherine and Apadula, Luca and Cepollaro, Carlo and Gomes, Henrique and Butterfield, Jeremy and Brukner, {\v{C}}aslav",
    title = "{Quantum coordinates, localisation of events, and the quantum hole argument}",
    doi = "10.1038/s42005-025-02084-3",
    journal = "Commun. Phys.",
    volume = "8",
    number = "1",
    pages = "185",
    year = "2025"
}

@article{tasakiLongrangeOrderTower2019,
  title = {Long-Range Order, ``Tower''' of States, and Symmetry Breaking in Lattice Quantum Systems},
  author = {Tasaki, Hal},
  year = {2019},
  journal = {Journal of Statistical Physics},
  volume = {174},
  number = {4},
  pages={735–761},
  issn = {0022-4715, 1572-9613},
  doi = {10.1007/s10955-018-2193-8}
}

@article{Kibble:1976sj,
    author = "Kibble, T. W. B.",
    title = "{Topology of Cosmic Domains and Strings}",
    reportNumber = "ICTP/75/5",
    doi = "10.1088/0305-4470/9/8/029",
    journal = "J. Phys. A",
    volume = "9",
    pages = "1387--1398",
    year = "1976"
}

@book{Cuffaro,
	editor = {Michael E. Cuffaro and Stephan Hartmann},
	publisher = {Oxford University Press},
	title = {Open Systems: Physics, Metaphysics, and Methodology},
	year = {2026}
}

@book{Georgii,
	address = {Berlin, New York},
	author = {Hans-Otto Georgii},
	edition = {2nd extended edition},
	publisher = {De Gruyter},
	title = {Gibbs Measures and Phase Transitions},
	year = {2011},
	doi = {10.1515/9783110250329},
	isbn = {9783110250329}
}

@article{Dobrushin1968,
  author  = {Dobrushin, R. L.},
  title   = {The Description of a Random Field by Means of Conditional Probabilities and Conditions of Its Regularity},
  journal = {Theory of Probability and Its Applications},
  volume  = {13},
  number  = {2},
  pages   = {197--224},
  year    = {1968},
  doi     = {10.1137/1113026}
}

@book{Tasaki2020,
  author    = {Tasaki, Hal},
  title     = {Physics and Mathematics of Quantum Many-Body Systems},
  series    = {Graduate Texts in Physics},
  publisher = {Springer Cham},
  year      = {2020},
  doi       = {10.1007/978-3-030-41265-4},
  isbn      = {978-3-030-41265-4}
}

@article{LanfordRuelle1969,
  author  = {Lanford, O. E. and Ruelle, D.},
  title   = {Observables at Infinity and States with Short Range Correlations in Statistical Mechanics},
  journal = {Communications in Mathematical Physics},
  volume  = {13},
  number  = {3},
  pages   = {194--215},
  year    = {1969},
  doi     = {10.1007/BF01645487}
}

@online{borsboominstantaneous,
	title = {Boundaries in the {Instantaneous} {Formulation} of {Field} {Theories}},
	author = {Borsboom, Silvester},
	year = {2026},
    note = {Forthcoming}
}

@book{Vilenkin:2000jqa,
    author = "Vilenkin, A. and Shellard, E. P. S.",
    title = "{Cosmic Strings and Other Topological Defects}",
    isbn = "978-0-521-65476-0",
    publisher = "Cambridge University Press",
    month = "7",
    year = "2000"
}

@article{horschSpincorrelationsLowLying1988,
  title = {Spin-Correlations and Low Lying Excited States of the Spin-1/2 Heisenberg Antiferromagnet on a Square Lattice},
  author = {Horsch, P. and Von Der Linden, W.},
  year = {1988},
  journal = {Zeitschrift für Physik B Condensed Matter},
  volume = {72},
  number = {2},
  issn = {0722-3277, 1431-584X},
  doi = {10.1007/BF01312134},
  pages = {181–193}
}

@article{Lessa_2025,
   title={Strong-to-Weak Spontaneous Symmetry Breaking in Mixed Quantum States},
   volume={6},
   ISSN={2691-3399},
   url={http://dx.doi.org/10.1103/PRXQuantum.6.010344},
   DOI={10.1103/prxquantum.6.010344},
   number={1},
   journal={PRX Quantum},
   publisher={American Physical Society (APS)},
   author={Lessa, Leonardo A. and Ma, Ruochen and Zhang, Jian-Hao and Bi, Zhen and Cheng, Meng and Wang, Chong},
   year={2025},
   month=Mar }

@article{hauser2026strongtoweaksymmetrybreakingopen,
    author = "Hauser, Jacob and Su, Kaixiang and Ha, Hyunsoo and Lloyd, Jerome and Kiely, Thomas G. and Vasseur, Romain and Gopalakrishnan, Sarang and Xu, Cenke and Fisher, Matthew P. A.",
    title = "{Strong-to-Weak Symmetry Breaking in Open Quantum Systems: From Discrete Particles to Continuum Hydrodynamics}",
    eprint = "2602.16045",
    archivePrefix = "arXiv",
    primaryClass = "quant-ph",
    month = "2",
    year = "2026"
}

@Article{10.21468/SciPostPhysLectNotes.24,
	title={{Phase transitions in the early universe}},
	author={Mark Hindmarsh and Marvin Lüben and Johannes Lumma and Martin Pauly},
	journal={SciPost Phys. Lect. Notes},
	pages={24},
	year={2021},
	publisher={SciPost},
	doi={10.21468/SciPostPhysLectNotes.24},
	url={https://scipost.org/10.21468/SciPostPhysLectNotes.24},
}

@book{vanwezel2026ssb,
    author = {Aron J. Beekman and Louk Rademaker and Jasper van Wezel},
    title = {Spontaneous Symmetry Breaking},
    publisher = {Cambridge University Press},
    year = {2026},
    note = {To be published},
    isbn = {9781009575768}
}

@article{Weir_2018,
   title={Gravitational waves from a first-order electroweak phase transition: a brief review},
   volume={376},
   ISSN={1471-2962},
   url={http://dx.doi.org/10.1098/rsta.2017.0126},
   DOI={10.1098/rsta.2017.0126},
   number={2114},
   journal={Philosophical Transactions of the Royal Society A: Mathematical, Physical and Engineering Sciences},
   publisher={The Royal Society},
   author={Weir, David J.},
   year={2018}
}

@article{schultz1964two,
  title={Two-dimensional Ising model as a soluble problem of many fermions},
  author={Schultz, Theodore D and Mattis, Daniel C and Lieb, Elliott H},
  journal={Reviews of Modern Physics},
  volume={36},
  number={3},
  pages={856},
  year={1964},
  publisher={APS},
  doi ={10.1103/RevModPhys.36.856}
}

\end{document}